\documentclass[pra,showpacs,nofootinbib]{revtex4-2}

\usepackage{hyperref}
\usepackage{enumitem}
\usepackage{amsmath}
\usepackage{amssymb}
\usepackage{amsthm}
\usepackage{graphics}
\usepackage{graphicx}
\usepackage{subcaption}
\usepackage{color}
\usepackage{mathrsfs}
\usepackage[usenames,dvipsnames,svgnames]{xcolor}
\usepackage{stmaryrd}

\long\def\ca#1\cb{}

\def\bra#1{\langle#1|}

\def\inpd#1#2{\langle#1|#2\rangle }
\def\ket#1{|#1\rangle }

\def\Tr#1{\textrm{Tr}\left(#1\right)}

\def\locc{\overline{\textrm{LOCC}}}  % closure of LOCC
\def\mmod#1{(\bmod{~#1})}
\def\myeq#1{Eq.~\eqref{#1}}
\def\AC{{\cal A}}
\def\BC{{\cal B}}
\def\CC{{\cal C}}

\def\HC{{\cal H}}
\def\IC{{\cal I}}

\def\MC{{\cal M}}

\def\OC{{\cal O}}

\def\RC{{\cal R}}

\def\XC{{\cal X}}

\def\ZC{{\cal Z}}
\def\endproof{{\hspace{\stretch{1}}$\blacksquare$}}

\newtheorem{thm1}{Theorem}
\newtheorem{obs1}{Observation}
\newtheorem{thm2}[thm1]{Theorem}
\newtheorem{thm3}[thm1]{Theorem}

\newtheorem{thm5}[thm1]{Theorem}
\newtheorem{thm6}[thm1]{Theorem}
\newtheorem{thm7}[thm1]{Theorem}

\newtheorem{thm9}[thm1]{Theorem}

\newtheorem{thm11}[thm1]{Theorem}

\newtheorem{lem1}{Lemma}
\newtheorem{cor1}{Corollary}

\newtheorem{lem2}[lem1]{Lemma}
\newtheorem{lem3}[lem1]{Lemma}
\newtheorem{lem4}[lem1]{Lemma}
\newtheorem{lem5}[lem1]{Lemma}
\newtheorem{lem6}[lem1]{Lemma}

\newtheorem{lem8}[lem1]{Lemma}

\newtheorem{lem10}[lem1]{Lemma}
\newtheorem{lem11}[lem1]{Lemma}
\newtheorem{lem12}[lem1]{Lemma}
\newtheorem{lem13}[lem1]{Lemma}
\newtheorem{lem14}[lem1]{Lemma}
\newtheorem{lem15}[lem1]{Lemma}
\newtheorem{lem16}[lem1]{Lemma}
\newtheorem{lem17}[lem1]{Lemma}
\newtheorem{lem18}[lem1]{Lemma}
\newtheorem{lem19}[lem1]{Lemma}
\newtheorem{lem20}[lem1]{Lemma}

\begin{document}
\title{Local approximation of multipartite quantum measurements}
\author{Scott M. Cohen}
%\ca
\email{cohensm52@gmail.com (he/him/his)}
\affiliation{Department of Physics, Portland State University, Portland Oregon 97201, USA}

\begin{abstract}
We provide a necessary condition that a quantum measurement can be implemented by the class of protocols known as Local Operations and Classical Communication, or LOCC, including when an error is allowed but must vanish in the limit of an infinite number of rounds, a case referred to as asymptotic LOCC. Our condition unifies, extends, and provides an intuitive, geometric justification for previous results on asymptotic LOCC. We use our condition to answer a variety of long-standing, unsolved problems, including for distinguishability of certain sets of states by LOCC. These include various classes of unextendible product bases, for which we prove they cannot be distinguished by LOCC even when infinite resources are available and asymptotically vanishing error is allowed.
\end{abstract}

\date{\today}
\pacs{03.65.Ta, 03.67.Ac}

\maketitle
\section{Introduction}
Beginning in the earliest days of quantum mechanics, even the leading scientists of the time struggled to understand the notion of quantum entanglement \cite{EPR,SchrodingerAll,Bohr1,Bohr2}, an unusual correlation that has no classical counterpart. More recently, entanglement has been found to underlie many of the key advancements in quantum information processing, including but not limited to, the surprising phenomenon of teleportation \cite{BennettTele,ExptTele6,ExptTele7}, quantum computing \cite{Benioff,Deutsch,FeynmanQComp,ShorFactor,MartinisQComp}, and quantum cryptography \cite{Ekert,BB84b}. Yet, our understanding of entanglement remains far from complete. In an effort to better understand this important phenomenon, researchers developed the first quantum resource theory \cite{ChitGourResource}, that for the resource of entanglement \cite{HoroRMP}. In general, such theories identify the states and operations that are available for free as separate from those that are, in some sense, costly. For entanglement, in particular, the free operations are known as \emph{local quantum operations and classical communication}, commonly denoted as LOCC. LOCC are the operations that cannot create entanglement between two or more spatially separated subsystems, and it has therefore long been recognized that an understanding of LOCC will greatly contribute to our understanding of entanglement (see, for example, Ref.~\cite{Nielsen,BKrausStateTransf}). Certainly, the study of LOCC has by now a long and distinguished history \cite{PeresWootters,BennettTele,Bennett9,BennettPurifyTele,Walgate}, driven also by the fact that LOCC has important practical applications in its own right \cite{CiracDistComp,BennettPurifyTele,Nielsen,BKrausStateTransf}. Nonetheless, limited progress has been made concerning the important asymptotic case of vanishingly small error \cite{Bennett9}, although recent years have seen renewed interest in this problem \cite{KKB,WinterLeung,FuLeungMancinska,ChitambarHsiehPeresWootters,ChildsLeung,ChitambarHsiehHierarchy}.

Here, we consider the problem of determining if a measurement ${\cal M}$ on a multipartite quantum system, consisting of any number of spatially separated parties $P$, can be carried out when those parties are constrained to the use of LOCC. In particular, our aim is to address ``asymptotic LOCC", here denoted $\locc$ (the topological closure of LOCC), for which an error is allowed in the implementation of ${\cal M}$, but that this error must become vanishingly small when the number of rounds used by the parties is allowed to grow without limit. Our main result, see Theorem~\ref{thm2} below, is a necessary condition that $\MC$ can be implemented by $\locc$. The key to our result is recognition of the significance of a geometric object $\ZC_\MC$, uniquely associated to $\MC$, followed by proof of existence of certain kinds of continuous paths lying within $\ZC_\MC$ when $\MC$ is in $\locc$. Theorem~\ref{thm2} subsumes previous results on $\locc$, which we will show are direct consequences of our theorem, and being geometric in nature, it provides important intuition as to when a given measurement is, or is not, in $\locc$. We here use Theorem~\ref{thm2} to answer a number of longstanding unsolved problems.

When measurement ${\cal M}$ is possible by $\locc$, which we will represent as $\MC\in\locc$, ${\cal M}$ can be approximated as closely as one wishes simply by including more and more rounds in the protocol. When this is not possible, then there is a non-zero gap between what can be accomplished by LOCC as compared to when ${\cal M}$ is implemented by global means. Several years ago, the authors in \cite{KKB} derived a necessary condition that ${\cal M}\in\locc$. Their condition was used to prove that when discriminating the ``double-trine" ensemble \cite{PeresWootters} of quantum states with minimum error, the optimal LOCC probability of success is strictly smaller than by global means \cite{ChitambarHsiehPeresWootters}. Their condition was also used to prove \cite{FuLeungMancinska} that a global measurement is strictly better than LOCC for discriminating any unextendible product basis \cite{IBM_CMP} on a $3\times3$ system. In addition, a ``nonlocality constant" $\eta$ is defined in \cite{ChildsLeung} and used to obtain a lower bound on the probability of error in LOCC discrimination of any set of bipartite states. A necessary condition for perfect discrimination by $\locc$ is then that $\eta=0$, a condition that is implied by the necessary condition of \cite{KKB}. These accomplishments notwithstanding, there remains a great deal to learn about asymptotic LOCC.

The remainder of the paper is organized as follows. We begin by reviewing how LOCC, and quantum measurements in general, can be mathematically described, and then we present our main result, Theorem~\ref{thm2}. In Section~\ref{sec3}, we provide a detailed proof for Theorem~\ref{thm2} and then in Section~\ref{sec4}, we illustrate its usefulness with several examples of local state discrimination, as well as by showing that the necessary conditions of \cite{KKB} (and therefore, \cite{ChildsLeung}), as well as \cite{ChitambarHsiehHierarchy}, all follow directly from our Theorem~\ref{thm2}. Finally, we close with a discussion of the implications of what we have found.

\section{Main Result}\label{sec2}
Any quantum measurement ${\cal M}$ may be thought of as a positive operator-valued measure, or POVM. A POVM consists of a set of operators, $E_j\ge0$, individually referred to as a POVM element. Each $E_j$ is a positive semidefinite operator, denoted as $E_j\ge0$, which means simply that the eigenvalues of $E_j$ are all non-negative. These operators satisfy a completeness relation,
\begin{align}\label{eqn100}
\sum_{j=1}^{|{\cal M}|}E_j={\cal I}_{\cal H},
\end{align}
where $|{\cal M}|$ is the number of POVM elements in ${\cal M}$, and ${\cal I}_{\cal H}$ is the identity operator on Hilbert space ${\cal H}$, of finite dimension $D$, describing states of the quantum system being measured.

An LOCC protocol implements an overall measurement ${\cal M}$ through a sequence of intermediate, local measurements by the individual parties. Such a protocol consists of one party making a measurement on their local system and then communicating the outcome of that measurement to the other parties. This is followed, according to a pre-approved plan, by the next party making a measurement and communicating the result to the others. Notice that only one party measures at a time, and they continue measuring and sharing classical information for however many rounds, possibly an infinite number, as is necessary. 

As is commonly done, we represent an LOCC protocol as a tree graph, consisting of nodes connected by an edge to each of its children. Each node $n$ is associated with a positive semidefinite operator $F_n$ representing the action of all parties up to that point in the protocol; a method for obtaining operators $F_n$ from the actual protocol is described in detail in the first paragraph of Section~II of \cite{myLOCCbyFirstMeas}. Because intermediate measurements are always local, each $F_n$ is of the tensor product form ${\cal A}_n\otimes{\cal B}_n\otimes\cdots$, where ${\cal A}_n$ is an operator on the first party's Hilbert space ${\cal H}_A$ and similarly for the other operators appearing in this expression.\footnote{Any positive semidefinite operator may be written as $F_n=f_n^\dagger f_n$, for some operator, $f_n$. That is, $F_n$ is Hermitian and acts as $F_n:\HC\to\HC$; the input and output spaces are the same. While the input and output spaces of $f_n$ may certainly differ from each other (and from $\HC$), this is not an issue of concern for us here, since we only need consider $F_n$. In any case, it is possible to show that for any protocol that involves intermediate maps $f_n$ that have input and output spaces that differ, there is an equivalent protocol with $f_n$ replaced by $f_n^\prime$, where $f_n^\prime:\HC\to\HC$, and then for each final outcome labeled by $l$, a single map (which will be an isometry when the output space is larger than the input) from $\HC$ to $\HC_l$ is implemented at the end of the protocol.} The root node represents the situation before any of the parties have done anything, and so is associated with the identity operator ${\cal I}_{\cal H}$. Every local measurement has multiple outcomes, each represented by one of the child nodes of their shared parent. Because this local measurement is complete, then in analogy to Eq.~\eqref{eqn100}, the sum of sibling child nodes is equal to their parent. A branch of the protocol begins at the root node and stretches from each node to one of its children, continuing without end in the case of an infinite branch, or terminating at what is known as a leaf node (leaf nodes being those that do not themselves have children). Those protocols that include infinite branches may be thought of as the limit of a sequence of finite-round protocols, and this is what is meant by asymptotic LOCC.

We will presently introduce certain \emph{convex sets}. Convex sets satisfy the condition that if $a$ and $b$ are each members of that set, then for any $0\le x\le1$, $(1-x)a+xb$ is also a member of that set. Starting with the convex set, $P$, consisting of all positive semidefinite operators, define the proper subset $\ZC\subset P$ as $\ZC=\left\{z\left\vert z=\sum_jc_jE_j, 0\le c_j\le1,E_j\in P\right.\right\}$. That is, $\ZC$ consists of all positive linear combinations of the operators, $E_j$, with coefficients not exceeding unity. Equivalently, $\ZC$ may be viewed as the Minkowski sum \cite{MinkowskiSum} of a set of line segments stretching from the zero operator to operators $E_j$. These line segments will be denoted as $[0,E_j]$, whereas the half-open line segment $(0,E_j]$ omits the zero operator. The geometric object $\ZC$, which is known as a \emph{zonotope}, may therefore also be written as $\ZC=\sum_j[0,E_j]$.

Consider an LOCC protocol implementing measurement $\MC$ consisting of POVM elements $E_j$. Let us envision each branch of this protocol as a path through $P$. Significantly, as we will see in the next section, these paths are confined within the zonotope, $\ZC\subset P$, defined above. We will refer to ``monotonic" paths, by which we mean a path of operators $\Pi(s)$ such that $\textrm{Tr}\left({\Pi(s)}\right)$ is a monotonic function of $s$. Then, we have our main theorem.
\begin{thm2}\label{thm2}
	If ${\cal M}\in\overline{\textrm{LOCC}}$, with measurement ${\cal M}$ consisting of POVM elements $E_j$, then for each $j$, there exists a continuous, monotonic path of product operators from ${\cal I}_{\cal H}$ to a point on the (half-open) line segment $(0,E_j]$, and this path lies entirely within zonotope $\ZC_{\cal M}=\sum_j[0,E_j]$.
\end{thm2}
The proof is given in the following section; those readers only interested in seeing how this theorem implies previously known conditions for asymptotic LOCC, as well as how it can be used to solve long outstanding problems, may skip ahead to Section~\ref{sec4}, where among other things, we will show that Theorem~\ref{thm2} implies the necessary conditions of \cite{KKB,ChildsLeung,ChitambarHsiehHierarchy}. In fact, we will show there that our theorem is strictly stronger than the first two of these conditions (and it is trivially stronger than the third, which applies only to cases of distinguishing a \emph{pair} of quantum states).

\section{Proof of Theorem~\ref{thm2}}\label{sec3}
In this section, we prove Theorem~\ref{thm2}. We follow the description given above of LOCC trees and how each node is associated with a positive operator representing the action of all parties up to that point in the protocol; see \cite{myLOCCbyFirstMeas} for more details. We will need Lemma~$1$ from \cite{myLOCCbyFirstMeas}, which is a straightforward consequence of the completeness of each intermediate measurement.
\begin{lem4}\cite{myLOCCbyFirstMeas}\label{lem4}
	Each node $n$ in a finite-round LOCC tree is equal to the sum of all leaf nodes that are descended from that node.
\end{lem4}

\noindent We will also use the following lemma in obtaining our main result.
\begin{lem1}\label{lem1}
	Given any measurement ${\cal M}$ with POVM elements $\{E_j\}$, the collection of operators $\ZC_{\cal M}:=\{\sum_{j}c_{j}E_j\vert 0\le c_j\le1~\forall{j}\}$ is a compact, convex set.
\end{lem1}
\proof Convexity is obvious. To see that it is compact, notice that $\ZC_{\cal M}=\sum_j[0,E_j]$, where this sum of line segments is of the Minkowski type, as discussed just above Theorem~\ref{thm2}. Each line segment is closed and bounded, hence compact, and the sum of compact sets is itself compact. Therefore, $\ZC_{\cal M}$ is compact.\hspace{\stretch{1}}$\blacksquare$

We choose as our metric on operator space to be the trace norm, $\left\|{X}\right\|=\textrm{Tr}\left({\sqrt{X^\dag X}}\right)$, which for positive semidefinite operators is equal to the trace. To prove our results, we also need to define a distance measure on POVMs. From a simplistic perspective, two POVMs will be identical when they share the same set of POVM elements. However, there may be cases when the number of elements in the two POVMs differ, so we need to take such a possibility into account. This will be important in studying LOCC, for which the number of outcomes grows increasingly larger with the number of rounds, $r$, even while the target measurement for that LOCC protocol may have a relatively small number of outcomes. For the LOCC and target POVMs to be identical, then, there would have to be many outcomes from the LOCC protocol that are the same, apart from a positive scalar factor. Thus, we recognize that for two POVMs to be identical, each outcome in the first POVM must be proportional to one of the outcomes in the second, and vice versa. In addition, identification of the two POVMs requires that the combined weights of the two sets of so identified elements must be equal. That is, if all proportional elements in the first POVM are added together to reduce each such subset to a single element, and the same is done for the second POVM, then there must be a one-to-one relationship between the elements of the two POVMs, each (reduced) element from the first being equal to the corresponding (reduced) element from the second. For simplicity in what follows, when we say measurement $\MC$ consists of outcomes $E_j$, we will assume such a reduction has already been carried out.

These considerations are precisely captured by the zonotopes discussed above \cite{AubrunLancienZonotopes}. For example, in direct analogy to the discussion above about combining weights of proportional POVM elements, the Minkowski sum of parallel line segments is just another parallel line segment having the combined length of the original subset. A POVM ${\cal M}$ with elements $E_j$ generates zonotope $\ZC_{\cal M}=\sum_j[0,E_j]$, and one can show that two zonotopes are identical if and only if their corresponding POVMs are identical in the sense discussed in the preceding paragraph. Therefore, we define our distance measure on POVMs to be the Hausdorff distance between the corresponding zonotopes. That is,
\begin{align}\label{eqn101}
	d({\cal M}_1,{\cal M}_2)=d_H(\ZC_1,\ZC_2)=\max\left\{\sup_{z_1\in \ZC_1}\inf_{z_2\in \ZC_2}\left\|{z_1-z_2}\right\|,\sup _{z_2\in \ZC_2}\inf_{z_1\in \ZC_1}\left\|{z_1-z_2}\right\|\right\}.
\end{align}

We follow \cite{WinterLeung} in drawing a distinction between two subsets of infinite-round LOCC, each of which may be discussed in terms of sequences of finite-round LOCC protocols. The first such subset involves infinite-round protocols that are the limit of sequences for which the next protocol in the sequence is the same as the previous one, except for the addition of one or more rounds at the end. The limit of such a sequence of protocols is in LOCC. By contrast, one may instead have a sequence of finite-round protocols for which the earlier rounds are allowed to change when adding rounds in going from one protocol to the next. Each of the protocols in the latter type of sequence implements an LOCC measurement, say ${\cal M}_r$, but the measurement ${\cal M}$ that is the asymptotic limit of this sequence of measurements may not itself be LOCC, instead only being in the topological closure, denoted as ${\cal M}\in\overline{\textrm{LOCC}}$.

We will show below that if measurement ${\cal M}\in\overline{\textrm{LOCC}}$, then for each $E_j\in{\cal M}$, there exists a particular set of monotonic paths of product operators lying entirely within the compact, convex set $\ZC_{\cal M}$ defined in Lemma~\ref{lem1}. 

We begin with the following observation, in which we introduce the concept of a ``piecewise-local path". Similar to that of a piecewise-constant curve, by this we mean a path made up of segments that are ``local", or in other words, segments for which one and only one of the parties' tensor parts is changing. This should be made clear by the examples given in the next paragraph.
\begin{obs1}\label{obs1}
	Each branch in an LOCC protocol, finite or infinite, constitutes a continuous, piecewise-local path in the convex set of positive semidefinite operators. Significantly, each point along this path is not just a positive semidefinite operator, but also a product operator of the form ${\cal A}\otimes{\cal B}\otimes{\cal C}\otimes\cdots$.
\end{obs1}
\noindent To see this, consider the sequence of positive operators labeling the nodes along a given branch, starting from the root node. Each such positive operator other than the root node represents the outcome of a local measurement by one of the parties. Being local, these measurements only change that particular party's part of the positive operator representing each outcome. Given that every protocol starts with a product operator, that being the identity operator, $I_A\otimes I_B\otimes\cdots$, then if Alice measures first with outcome ${\cal A}$, a continuous path of product operators from the identity operator to that outcome can be parameterized by $x$ ranging from $0$ to $1$ as $[(1-x)I_A+x{\cal A}]\otimes I_B\otimes\cdots$. Given that it is only the operator on ${\cal H}_A$ that changes, this piece of the path is local. If Bob measures next with outcome ${\cal B}$, this path is similarly parametrized as ${\cal A}\otimes[(1-y)I_B+y{\cal B}]\otimes\cdots$, which is another local piece for which only the ${\cal H}_B$ part of the operator changes. The remainder of the continuous path may be generated in the same fashion for each and every branch in the protocol, and it is clear that these paths consist of pieces that are local, as claimed. In the limit of infinite-round LOCC protocols, these paths still exist, just now with an infinite number of piecewise-local segments.

As a consequence of this observation, we obtain the following theorem, which provides a necessary condition for LOCC.
\begin{thm1}\label{thm1}
	If an LOCC protocol, finite or infinite, implements a measurement ${\cal M}$ consisting of outcomes $E_j$, then for each $j$, there exists at least one continuous, monotonic, piecewise-local path from $I_{\cal H}$ to $(0,E_j]$, and every point along that path is a product operator. In addition, each of these paths lies entirely within the zonotope, $\ZC_{\cal M}=\sum_j[0,E_j]$.
\end{thm1}
\proof The proof was given above, apart from the claims that the path is monotonic, terminates on $(0,E_j]$, and lies in $\ZC_{\cal M}$. Monotonicity follows immediately from the recognition that each child node represents one of (generally) multiple outcomes of a measurement made at the parent node. Since the sum of the children, say $F_n$, is equal to their parent, $F_p$, then $\textrm{Tr}\left({F_p}\right)=\sum_n\textrm{Tr}\left({F_n}\right)\ge\textrm{Tr}\left({F_m}\right)$, for any $F_m$ in the set of children. Monotonicity is then evident from the fact that our paths proceed from parent to child, child to grandchild, and so on.

To show for finite protocols that the path terminates on $(0,E_j]$, notice that every leaf node terminating a branch is proportional to one of the $E_j$, since otherwise the protocol does not implement ${\cal M}$. This implies that each leaf $l$ is $\hat E_l=q_lE_j$, for some $j$ and $0<q_l\le1$ ($q_l$ cannot exceed unity because to implement $\MC$, the sum of all leaf nodes proportional to $E_j$ must equal $E_j$). The argument just below Observation~\ref{obs1} indicates that the path terminates at $\hat E_l$, which lies on $(0,E_j]$, as claimed. To show that this path lies in $\ZC_{\cal M}$, also notice that every node in the finite LOCC tree is a sum of the leaf nodes that descend from that node, see Lemma~\ref{lem4} above. It then follows immediately that every node in the tree is an element of $\ZC_{\cal M}$, and since every point on the considered path is a convex combination of a pair of nodes in the tree---those nodes being the ancestor and descendant that are nearest to the point in question, see the explanation following Observation~\ref{obs1}---these points also lie in $\ZC_\MC$, and this completes the proof for the finite case.

For infinite protocols in LOCC, recall the discussion above that these are the limit of sequences of protocols that arise by simply adding one or more additional rounds at the end of branches present in the previous protocol of the sequence. Therefore in the limit, each branch generates a path as described in the theorem, but in this case some of those branches become infinite in the limit. While these paths now have an infinite number of steps, all those steps continue to be piecewise-local in the limit. As just discussed, for finite protocol ${\cal P}_r$, the corresponding path lies entirely within zonotope $\ZC_{{\cal M}_r}$. In addition, each leaf node in ${\cal P}_r$ that is not also a leaf node in ${\cal P}_{r+1}$ is followed (and still present, though no longer a leaf) in ${\cal P}_{r+1}$ by one more complete local measurement. Since the children produced by that one extra measurement sum to their parent, it is easy to see that $\ZC_{{\cal M}_r}\subseteq \ZC_{{\cal M}_{r+1}}$. Thus, the piecewise-local paths generated by ${\cal P}_r$ not only lie entirely within $\ZC_{{\cal M}_r}$, but also within $\ZC_{{\cal M}_{r^\prime}}$ for all $r^\prime\ge r$. Taking the limit, we have that $\ZC_{\cal M}=\lim_{r\to\infty}\ZC_{{\cal M}_r}$ (since by assumption, the infinite-round protocol implements ${\cal M}$), $\ZC_{{\cal M}_r}\subseteq \ZC_{\cal M}$ for all $r$, and it follows that the path to each of the outcomes of ${\cal P}=\lim_{r\to\infty}{\cal P}_r$ lies entirely within $\ZC_{\cal M}$. This completes the proof.\hspace{\stretch{1}}$\blacksquare$

\noindent Note that monotonicity is important because it excludes the trivial path, which exists for any measurement whatsoever, along $s{\cal I}_{\cal H}$ from ${\cal I}_{\cal H}$ to $0$ and then along $sE_j$ from $0$ to a point on the line segment $(0,E_j]$.

By dropping the condition that the paths be piecewise-local, we can now prove Theorem~\ref{thm2}.

\noindent\emph{Proof of Theorem~\ref{thm2}}. The condition of this theorem, ${\cal M}\in\overline{\textrm{LOCC}}$, means there exists a sequence of finite-round LOCC measurements $\{{\cal M}_r\}$ such that $\lim_{r\to\infty}{\cal M}_r={\cal M}$. This implies that for all $\epsilon>0$ there exists $R\in\mathbb{N}$ such that for all $r>R$,
\begin{align}\label{eqn1014}
	\epsilon>d\left({\cal M},{\cal M}_r\right)=d_H(\ZC_{\cal M},\ZC_{{\cal M}_r}),
\end{align}
and therefore, $\ZC_{\cal M}=\lim_{r\to\infty}\ZC_{{\cal M}_r}$. We wish to show that there exists a continuous, monotonic path of product operators from $I_{\cal H}$ to $(0,E_j]$, for each $j$, and that these paths lie in $\ZC_{\cal M}$. By Theorem~\ref{thm1}, for each leaf node $\hat E_l^{(r)}$ of protocol $P_r$ implementing ${\cal M}_r$, we know there is a path of product operators from $I_{\cal H}$ to $\hat E_l^{(r)}$ lying entirely within $\ZC_{{\cal M}_r}$.\footnote{Note, however, that $\ZC_{{\cal M}_r}\subseteq \ZC_{{\cal M}_{r+1}}$ need not hold here because we allow earlier rounds to change in going from one protocol in the sequence to the next. Note also that, according to the discussion following Observation~\ref{obs1} and the proof of Theorem~\ref{thm1} for the finite case, this path terminates not just along $(0,\hat E_l^{(r)}]$, but precisely at $\hat E_l^{(r)}$.} Thus, there exists a sequence of continuous, monotonic paths of product operators $\{\Pi_r(s)\}_r$ from ${\cal I}_{\cal H}$ to each outcome $\hat E_l^{(r)}$ of ${\cal P}_r$, each path lying in $\ZC_{{\cal M}_r}$. The condition $\lim_{r\to\infty}{\cal M}_r={\cal M}$ means that for each $E_j\in{\cal M}$ there exists a sequence of indices, $\{l_r\}_r$, such that $\lim_{r\to\infty}\hat E_{l_r}^{(r)}=qE_j$ for some $0<q\le1$ (generally, many such sequences will exist for each $j$). This sequence of indices therefore corresponds to a sequence ${\cal S}_j$ of paths whose endpoints $\hat E_l^{(r)}$ converge to a point on $(0,E_j]$ as $r\to\infty$. Starting with ${\cal S}_j$, we can now show the existence of a continuous, monotonic path of product operators $\Pi(s)\in \ZC_{\cal M}$ from ${\cal I}_{\cal H}$ to a point on $(0,E_j]$. Since $j$ is arbitrary here, this conclusion will then hold for each $j$.

Path $\Pi_r(s)$ consists of a sequence of piecewise-local segments starting from $F_0={\cal I}_{\cal H}$ at node $n=0$ and continuing with each local measurement to node $n=1,2,\cdots,r^\prime$ (we include $r^\prime\le r$ to allow for branches that terminate before round $r$). The $(n+1)$th segment of this path starts at parent node $F_n$ and follows a straight line to child node $F_{n+1}$. Let us parametrize this path in terms of the trace---with $s$ decreasing through the interval $\left[0,D\right]$ starting at $D$ and moving toward $0$. Then, each segment is
\begin{align}\label{eqn202}
	\Pi_r(s)=\frac{\left[s-\textrm{Tr}\left({F_{n+1}}\right)\right]F_n+\left[\textrm{Tr}\left({F_n}\right)-s\right]F_{n+1}}{\textrm{Tr}\left({F_n}\right)-\textrm{Tr}\left({F_{n+1}}\right)},~~\textrm{Tr}\left({F_n}\right)\ge s\ge\textrm{Tr}\left({F_{n+1}}\right)
\end{align}
for each node starting at $n=0$ and continuing to $n=r^\prime-1$. We add a constant final piece to this path, $\Pi_r(s)=\hat E_l^{(r)}$ for $\textrm{Tr}\left({\hat E_l^{(r)}}\right)\ge s\ge0$, so that $\Pi_r(s)$ is defined over the full interval $s\in[0,D]$. Taking the trace of \eqref{eqn202}, we see that $s=\textrm{Tr}\left({\Pi_r(s)}\right)$ for $D\ge s\ge\textrm{Tr}\left({\hat E_l^{(r)}}\right)$.

Next, we show that for each $r$, $\Pi_r(s)$ is Lipschitz continuous, which means that $\left\|{\Pi_r(s)-\Pi_r(s^\prime)}\right\|\le K\left\vert{s-s^\prime}\right\vert$ for some constant $K$, referred to as the Lipschitz constant. For $\textrm{Tr}\left({\hat E_l^{(r)}}\right)\le s^\prime\le s$, we have 
\begin{align}\label{eqn204}
	s-s^\prime&=\textrm{Tr}\left({\Pi_r(s)-\Pi_r(s^\prime)}\right)\notag\\
	&=\left\|{\Pi_r(s)-\Pi_r(s^\prime)}\right\|,
\end{align}
which follows for the trace norm because, as is easily seen from how we've constructed these paths, $\Pi_r(s)-\Pi_r(s^\prime)$ is positive semidefinite, and so has non-negative eigenvalues. In addition, for $s^\prime\le s\le\textrm{Tr}\left({\hat E_l^{(r)}}\right)$, $\Pi_r(s)=\Pi_r(s^\prime)$; while for $s^\prime\le\textrm{Tr}\left({\hat E_l^{(r)}}\right)\le s$, $s-s^\prime\ge\textrm{Tr}\left({\Pi_r(s)-\hat E_l^{(r)}}\right)=\textrm{Tr}\left({\Pi_r(s)-\Pi_r(s^\prime)}\right)=\left\|{\Pi_r(s)-\Pi_r(s^\prime)}\right\|$. Thus, $\Pi_r(s)$ is Lipschitz continuous over the entire interval $s\in[0,D]$, as claimed, with Lipschitz constant $K=1$, independent of $r$.

Therefore, we may apply the Arzel\`a-Ascoli theorem \cite{ArzelaAscoli}, which tells us that for any sequence of Lipschitz continuous paths for which all instances share the same Lipschitz constant, there exists a subsequence that converges uniformly on $[0,D]$ to a continuous limiting path, $\Pi(s)$. In addition, $\Pi(s)$ is also Lipschitz continuous with the same Lipschitz constant. For each $s$, $\Pi(s)$ is thus a limit of points $\Pi_r(s)$ in this subsequence, implying that $\Pi(s)$ is arbitrarily close to a product operator, and is therefore itself a product operator.\footnote{This is easily proved using the Eckart–Young–Mirsky theorem.\cite{EckartYoungMirsky}} That is, $\Pi(s)$ is a continuous path of product operators. Furthermore, since (1) every path in the subsequence is monotonic and lies within its corresponding zonotope $\ZC_{{\cal M}_r}$; (2) $\lim_{r\to\infty}\ZC_{{\cal M}_r}=\ZC_{\cal M}$; and (3) $\ZC_{\cal M}$ is a compact set, see Lemma~\ref{lem1}; then $\Pi(s)$ is also monotonic and lies in $\ZC_{\cal M}$. Finally, since we started with a sequence ${\cal S}_j$ of paths whose endpoints converge to a point on $(0,E_j]$, the chosen subsequence must also have endpoints converging to that same point on $(0,E_j]$, and this completes the proof.\hspace{\stretch{1}}$\blacksquare$

Theorem~\ref{thm2} provides a powerful method for the study of approximate implementation of quantum measurements by LOCC. Consider the local state discrimination problem \cite{Dieks,Walgate,WalgateHardy,Chefles,ChildsLeung,KKB,Hayashi,myLDUPB}, wherein a referee prepares a multipartite system in one of a known set of states and distributes the subsystems to the individual parties, who are then tasked with determining in which one of the states the system was prepared. In the following section, we use Theorem~\ref{thm2} to show for several long-standing unsolved examples of local state discrimination, including in several cases of unextendible product bases \cite{IBM_CMP}, that the parties cannot accomplish the given task with arbitrarily small error when restricted to using LOCC. For completeness, we also show this holds for a few cases that were previously known, often accomplishing our proof in a simpler, more direct way. These examples include cases of perfect discrimination, where the parties always know with certainty which state was prepared, as well as a class of unambiguous state discrimination problems, see Section~\ref{subsecG}. In addition, we use Theorem~\ref{thm2} to show that the POVM constructed in \cite{ChitambarHsiehPeresWootters} for optimal discrimination of the double-trine states, initially studied in the seminal paper of Peres and Wootters \cite{PeresWootters}, cannot be implemented in $\overline{\textrm{LOCC}}$, a result also obtained in \cite{ChitambarHsiehPeresWootters}. Often one discovers that for these measurements, $(0,{\cal I}_{\cal H}]$ is an isolated line segment in the intersection of the set of product operators with $\ZC_{\cal M}$, which according to the following corollary to Theorem~\ref{thm2}, shows directly that ${\cal M}\not\in\overline{\textrm{LOCC}}$.
\begin{cor1}\label{cor1}
	Given measurement ${\cal M}$, consisting of POVM elements $E_j$, then $\MC\not\in\locc$ if any one or more of the following are isolated in the intersection of $\ZC_\MC$ with the set of (non-zero) product operators: $(a)$  $(0,E_j]$, for any $j$; or $(b)$ $(0,I_\HC]$.
\end{cor1}
\proof Case $(a)$ is obvious, given Theorem~\ref{thm2}. For case $(b)$, simply note that when $(0,I_\HC]$ is isolated, then the only way to get from $I_\HC$ to $(0,E_j]$ along a continuous path of product operators is to go directly along a line from $I_\HC$ to $0$, and then from $0$ back out along the line segment $(0,E_j]$. However, this is not monotonic, so does not satisfy the conditions of Theorem~\ref{thm2}.\hspace{\stretch{1}}$\blacksquare$

\noindent We will use this corollary in the next section to prove that certain sets of states cannot be distinguished within $\locc$.

\section{Applications}\label{sec4}
In this section, we illustrate the power of Theorem~\ref{thm2} by applying it to a series of examples.
\subsection{Understanding Proposition~$1$ of \cite{KKB}}
As our first illustration of the power of our theorem, we show below that Proposition~$1$ of \cite{KKB} is a direct consequence of Theorem~\ref{thm2}, lending insight into the meaning of Proposition~$1$. We then show that our theorem is able to answer questions their proposition is incapable of answering. Specifically, for the set of states discussed in \cite{KKB}, which they show that their Proposition $1$ \emph{cannot} determine whether there exists a measurement ${\cal M}\in\overline{\textrm{LOCC}}$ that accomplishes perfect state discrimination of the set, we will use Theorem~\ref{thm2} to answer this question in the negative. First, we restate Proposition~$1$ of \cite{KKB}. 

{\bf Proposition $1$.}\cite{KKB} \textit{Let $\{\rho_\mu\}$ be a family of $N$ states, such that $\cap_\mu\ker{(\rho_\mu)}$ contains no product vector (except $0$). Then $\{\rho_\mu\}$ can be discriminated perfectly by asymptotic LOCC only if for all $\chi$ with $1/N\le\chi\le1$ there exists a product operator $R\ge0$ obeying $\sum_\mu \textrm{Tr}\left({R\rho_\mu}\right)=1$, $\max_\mu Tr(R\rho_\mu)=\chi$, and $Tr(R\rho_\mu R\rho_\nu)=0$ for $\mu\ne\nu$.}

\noindent We will now see that the existence of product operator $R$ of this proposition for the given continuous range of $\chi$ follows directly from the continuous paths of product operators required by our Theorem~\ref{thm2}, as is demonstrated by the proof of the following lemma.
\begin{lem5}\label{lem5}
	Proposition 1 of \cite{KKB} is a direct consequence of our Theorem~\ref{thm2}.
\end{lem5}
\proof Consider measurement ${\cal M}$ consisting of outcomes $E_j$ and partition the outcomes into distinct sets $J_\mu$ such that $\textrm{Tr}\left({E_j\rho_\mu}\right)=0$ for all $j\not\in J_\mu$. Such a partition must be possible for perfect discrimination of these states by ${\cal M}$, since otherwise there will be errors. For $0\le s\le1$, define
\begin{align}\label{eqn2001}
	R(s)=\frac{\Pi(s)}{\sum_\mu\textrm{Tr}\left({\Pi(s)\rho_\mu}\right)},
\end{align}
where $\Pi(s)=\sum_jc_j(s)E_j$, $0\le c_j(s)\le1$, is a continuous path of positive semidefinite product operators, guaranteed to exist by Theorem~\ref{thm2}, from ${\cal I}_{\cal H}$ to $(0,E_{\hat\jmath}]$, for the specific outcome $\hat\jmath$. As such, $R(s)$ is a continuous function of $s$. We set $c_j(0)=1$ for all $j$ so that $\Pi(0)={\cal I}_{\cal H}$ and $R(0)={\cal I}_{\cal H}/N$, and $c_j(1)=0$ for all $j\ne\hat\jmath$, so that $\Pi(1)\propto E_{\hat\jmath}$ and $R(1)=E_{\hat\jmath}/\textrm{Tr}\left({E_{\hat\jmath}\rho_{\hat\mu}}\right)$, where $\hat\jmath\in J_{\hat\mu}$. Note that the condition of Proposition $1$, that $\cap_\mu\ker{\rho_\mu}$ contains no nonzero product operator, is necessary to ensure the denominator of $R(s)$ does not vanish.

We have defined $R(s)\ge0$ such that $\sum_\mu\textrm{Tr}\left({R(s)\rho_\mu}\right)=1$ for all $s$. Since $E_j\ge0$ and $\rho_\mu\ge0$, we have that $\textrm{Tr}\left({E_j\rho_\mu}\right)=0\implies E_j\rho_\mu=0=\rho_\mu E_j$ for all $j\not\in J_\mu$, and we see immediately that $\textrm{Tr}\left({R(s)\rho_\mu R(s)\rho_\nu}\right)=0$ for all $s$ and for all $\mu\ne\nu$. Finally, we must show that there exists $s$ such that $f(s)\equiv\max_\mu{\textrm{Tr}\left({R(s)\rho_\mu}\right)}=\chi$ for all $\chi$ in the range $1/N\le\chi\le1$. First, note that $f(0)=1/N$ and $f(1)=1$, each of which are easily seen from the expressions for $R(0)$ and $R(1)$ given in the preceding paragraph. The result for all $\chi$ then follows immediately due to continuity of $f(s)$ for $0\le s\le1$, which itself follows from continuity of $R(s)$. This ends the proof.\hspace{\stretch{1}}$\blacksquare$

Next, we show that Theorem~\ref{thm2} is, in fact, strictly stronger than Proposition $1$ of \cite{KKB}, by using our theorem to demonstrate the fact (which cannot be shown by Proposition $1$ \cite{KKB}) that no measurement ${\cal M}\in\overline{\textrm{LOCC}}$ accomplishes perfect state discrimination of the set of mutually orthogonal states given in their Eq.~(15). We will see that there is only one measurement ${\cal M}$ that works for this task, and that the requisite (by Theorem~\ref{thm2}) path of product operators in $\ZC_{\cal M}$ is non-existent for at least one of the outcomes in ${\cal M}$.

The (pairwise orthogonal) states in the set Eq.~(15) of \cite{KKB} are
\begin{align}\label{eqn1007}
	\vert{\psi_{1}}\rangle&\propto\vert{00}\rangle\notag\\
	\vert{\psi_{2}}\rangle&\propto2\vert{01}\rangle-\left(\sqrt{3}+1\right)\vert{10}\rangle-\sqrt{6}\sqrt[\leftroot{0}\uproot{0}4]{3}\vert{11}\rangle\notag\\
	\vert{\psi_{3}}\rangle&\propto2\vert{01}\rangle-\left(\sqrt{3}-1\right)\vert{10}\rangle+\sqrt{2}\sqrt[\leftroot{0}\uproot{0}4]{3}\vert{11}\rangle.
\end{align}
The only state orthogonal to all of these is proportional to
\begin{align}\label{eqn1008}
	\vert{\phi}\rangle&\propto2\sqrt[\leftroot{0}\uproot{0}4]{3}\vert{01}\rangle+\sqrt[\leftroot{0}\uproot{0}4]{3}\left(\sqrt{3}+1\right)\vert{10}\rangle-\sqrt{2}\vert{11}\rangle.
\end{align}
Therefore, the only (refined) measurement discriminating these states must consist of elements that project onto states that are superpositions of $\vert{\phi}\rangle$ with one (and then being orthogonal to the other two) of the states in Eq.~\eqref{eqn1007}.
There are six such states that are product, as required, including $\vert{\psi_{1}}\rangle$. They are
\begin{align}\label{eqn1009}
	\vert{\psi_{11}}\rangle&\propto\vert{\psi_{1}}\rangle=\vert{0}\rangle\otimes\vert{0}\rangle\notag\\
	\vert{\psi_{12}}\rangle&\propto\left(\sqrt{2}\sqrt[\leftroot{0}\uproot{0}4]{3}\vert{0}\rangle-\vert{1}\rangle\right)\otimes\left[\sqrt[\leftroot{0}\uproot{0}4]{3}\left(\sqrt{3}+1\right)\vert{0}\rangle-\sqrt{2}\vert{1}\rangle\right]\notag\\
	\vert{\psi_{21}}\rangle&\propto\left(\sqrt[\leftroot{0}\uproot{0}4]{3}\vert{0}\rangle-\sqrt{2}\vert{1}\rangle\right)\otimes\vert{1}\rangle\notag\\
	\vert{\psi_{22}}\rangle&\propto\vert{1}\rangle\otimes\left[\sqrt[\leftroot{0}\uproot{0}4]{3}\left(\sqrt{3}+1\right)\vert{0}\rangle+\sqrt{2}\vert{1}\rangle\right]\notag\\
	\vert{\psi_{31}}\rangle&\propto\left(3^{3/4}\vert{0}\rangle+\sqrt{2}\vert{1}\rangle\right)\otimes\vert{1}\rangle\notag\\
	\vert{\psi_{32}}\rangle&\propto\vert{1}\rangle\otimes\left[3^{3/4}\sqrt{2}\vert{0}\rangle-\left(\sqrt{3}+1\right)\vert{1}\rangle\right]
\end{align}
Define $\psi_{ij}=\ket{\psi_{ij}}\bra{\psi_{ij}}$. Allowing for higher-rank measurements, any outcome that identifies $\ket{\psi_i}$ with certainty must be a mixture of the form, $c_{i1}\psi_{i1}+c_{i2}\psi_{i2}$, but these are not product operators (as required for $\locc$) unless $c_{i1}c_{i2}=0$, so the measurement must be rank-$1$ with each outcome proportional to one of the $\psi_{ij}$. From the condition that the measurement is complete, \myeq{eqn100}, we find that $\psi_{12}$ must be excluded. There is one (and only one) complete measurement ${\cal M}$ consisting of elements proportional to the remaining five projectors. We will show that within the intersection of the set of product operators on ${\cal H}$ with the zonotope $\ZC_{\cal M}$ generated by these five projectors, $(0,\psi_{11}]$ is isolated, so that no continuous, monotonic path of product operators exists in $Z_{\cal M}$ from ${\cal I}_{\cal H}$ to this segment. By Theorem~\ref{thm2}, we then have that ${\cal M}\not\in\overline{\textrm{LOCC}}$, and the states of \myeq{eqn1007} cannot be discriminated within asymptotic LOCC.

Consider all product operators in $\ZC_{\cal M}$, which are of the form
\begin{align}\label{eqn1010}
	\AC\otimes\BC=c_{11}\psi_{11}+\sum_{k=2}^3\sum_{j=1}^2c_{kj}\psi_{kj}.
\end{align}
By writing the right-hand side out explicitly as a $4\times4$ matrix, and noting that in order to be a product operator the four $2\times2$ blocks must be proportional to each other, one finds that (1) if ${\cal B}$ is not diagonal then ${\cal A}={\cal A}_{11}[1]$, which is diagonal ($[j]=\vert{j}\rangle\langle{j}\vert,~j=0,1$); and (2) if ${\cal A}$ is not diagonal, then ${\cal B}={\cal B}_{11}[1]$, which is also diagonal. Significantly for our purposes, neither of these cases is anywhere near being proportional to $\psi_{11}=[0]\otimes[0]$. Therefore, if $\AC\otimes\BC$ is to be close to $(0,\psi_{11}]$, then ${\cal A}$ and ${\cal B}$ are both diagonal, from which one finds that $c_{2j}=\sqrt{3}c_{3j}$ for $j=1,2$, and $\AC\otimes\BC$ reduces to
\begin{align}\label{eqn1012}
	{\cal A}\otimes{\cal B}&=(1+\sqrt{3})\{c_{11}^\prime [00]+3c_{31}[01]\notag\\
	&+12c_{32}[10]+2(c_{31}+2c_{32})[11]\},
\end{align}
with $c_{11}=(1+\sqrt{3})c_{11}^\prime$. The right-hand side must be a product operator, implying
\begin{align}\label{eqn1013}
	(3c_{31})(12c_{32})&=2\left(c_{31}+2c_{32}\right)c_{11}^\prime\notag\\
	&\ge4c_{11}^\prime c_{32}.
\end{align}
This implies that either $c_{32}=0$ or $c_{31}\ge c_{11}^\prime/9$. A completely analogous argument shows that either $c_{31}=0$ or $c_{32}\ge c_{11}^\prime/18$. Note that if $c_{31}=0\ne c_{32}$ or $c_{31}\ne0=c_{32}$, then $c_{11}^\prime=0$, a case we may exclude because  we are seeking operators near $(0,\psi_{11}]$. 
Therefore, Eq.~\eqref{eqn1013} implies either (i) $c_{31}=0=c_{32}$, which leaves us with $c_{11}\psi_{11}\in(0,\psi_{11}]$; or (ii) $c_{32}\ge c_{11}^\prime/18$ and $c_{31}\ge c_{11}^\prime/9$, which is not near $(0,\psi_{11}]$. Hence, the only product operators near $(0,\psi_{11}]$ are proportional to $\psi_{11}$ itself, and we thus see that $(0,\psi_{11}]$ is an isolated line segment in the intersection of the set of product operators with $\ZC_{\cal M}$ as claimed, and this concludes the proof.\footnote{Let us note that function $E_\chi$, found in the appendix of \cite{KKB}, satisfies the conditions of their Proposition $1$ and is of a form very similar to that given here in Eq.~\eqref{eqn2001}. However, while $E_\chi$ provides a path of product operators from ${\cal I}_{\cal H}$ to $(0,\psi_{31}]$ lying in $\ZC_{\cal M}$ in its entirety, it is not continuous at $\chi=1/2$. In any case, it does not approach $(0,\psi_{11}]$, which we have just shown is impossible.}

\subsection{The necessary condition of \cite{ChitambarHsiehHierarchy}}\label{sec0}
Here, we use our Theorem~\ref{thm2} to derive the necessary condition of \cite{ChitambarHsiehHierarchy} that a pair of states can be perfectly discriminated by asymptotic LOCC, see their Theorem~$1$, reproduced here.
\begin{thm3}\label{thm3}\cite{ChitambarHsiehHierarchy}
	If $N$-partite states $\rho$ and $\sigma$ can be perfectly distinguished by asymptotic LOCC,
	then for each $x\in[1/2, 1]$ there must exist a POVM $\{\Pi_0,\Pi_\lambda\}_{\lambda=1}^D$ such that $\Pi_0$ is a separable operator, each $\Pi_\lambda$ is a product operator, and
	\begin{align}
		\Tr{\Pi_0\rho} &= 0,\label{eqn1101}\\
		\Tr{\Pi_\lambda\rho\Pi_\lambda\sigma} &= 0, ~~~\forall{1\le\lambda\le D},\label{eqn1002}\\
		\Tr{\Pi_\lambda[(1-x)\rho -x\sigma]} &= 0, ~~~\forall{1\le\lambda\le D}, \label{eqn1003}
		`	\end{align}
	where $D=\prod\limits_{k=1}^Nd_k^2+1$ and $d_k$ is the dimension of system $k$.
\end{thm3}
Let us now show that Theorem~\ref{thm3} follows from our necessary condition, Theorem~\ref{thm2}, that there exists a continuous path of product operators from $\IC_\HC$ to $(0,E_j]$ for each $j$. Here, $\{E_j\}$ is a POVM $\MC$ that perfectly distinguishes $\{\rho,\sigma\}$, so that there exists a partition of $\MC$ into those elements $j\in S_\rho$ that identify $\rho$ and those elements $j\in S_\sigma$ identifying $\sigma$. Then, $\Tr{E_j\rho}=0$ for $j\in S_\sigma$ and $\Tr{E_j\sigma}=0$ for $j\in S_\rho$. Define our continuous path of product operators lying in $\ZC_\MC$  from $\IC_\HC$ to $(0,E_{\hat\jmath}]$ for $\hat\jmath\in S_\rho$ as
\begin{align}\label{eqn1004}
	R_{\hat\jmath}(s)=\sum_jc_j^{(\hat\jmath)}(s)E_j,
\end{align}
where $0\le s\le1$, $0\le c_j^{(\hat\jmath)}(s)\le1$, $c_j^{(\hat\jmath)}(0)=1$ for all $j$, and $c_j^{(\hat\jmath)}(1)=\delta_{j\hat\jmath}$. This path is guaranteed to exist by our Theorem~\ref{thm2}.

Set $\Pi_0=\sum_{j\in S_\sigma}E_j$, which is manifestly separable, and satisfies \myeq{eqn1101}, as required for Theorem~\ref{thm3}. Note that $\Tr{E_i\rho E_j\sigma}=0$ for all $i,j\in S_\rho$, because $\Tr{X\sigma}=0$ implies $X\sigma=0=\sigma X$ when $X\ge0$ (since $\sigma\ge0$). Therefore, the replacement $\Pi_\lambda=R_{\hat\jmath}(s)$ satisfies \myeq{eqn1002}. It also satisfies \myeq{eqn1003} for each $x\in[1/2,1]$, as we now demonstrate. Solving for $x$, we find
\begin{align}\label{eqn1005}
	x(s)&=\frac{\Tr{R_{\hat\jmath}(s)\rho}}{\Tr{R_{\hat\jmath}(s)\rho}+\Tr{R_{\hat\jmath}(s)\sigma}}\notag\\
	&=\frac{\sum_jc_j^{(\hat\jmath)}(s)\Tr{E_j\rho}}{\sum_j\left[c_j^{(\hat\jmath)}(s)\Tr{E_j\rho}+c_j^{(\hat\jmath)}(s)\Tr{E_j\sigma}\right]}
\end{align}
Each term in the sum in the denominator is non-negative and at least some terms do not vanish. Thus, the denominator is nonzero. We know that $\sum_jc_j^{(\hat\jmath)}(s)E_j$ is continuous in $s$, and therefore $x=x(s)$ is a continuous function of $s$. For $s=0$, $c_j^{(\hat\jmath)}(0)=1$ for all $j$ so that $R_{\hat\jmath}(s)=\IC_\HC$, and we see that $x(0)=1/2$.  At the other end of the range of $s$ we have that $c_j^{(\hat\jmath)}(1)=\delta_{j\hat\jmath}$ so that $R_{\hat\jmath}(1)=E_{\hat\jmath}$. Recalling the choice $\hat\jmath\in S_\rho$ so that $\Tr{E_{\hat\jmath}\sigma}=0$, we have that $x(1)=1$. By continuity, then, $x(s)$ takes on all values between $1/2$ and $1$, and the conditions of the theorem are satisfied by replacing $\Pi_\lambda\to R_{\hat\jmath}(s)$ for any $\hat\jmath$. In the case where $\vert S_\rho\vert<D$, there are not enough values of $\hat\jmath$ available, but we can simply add zero operators to make up the difference. On the other hand, when $\vert S_\rho\vert>D$, we can reduce the number by using the same argument used in \cite{ChitambarHsiehHierarchy}, by way of Carath\'eodory's Theorem. Thus, we see that our Theorem~\ref{thm2}, guaranteeing a continuous path of product operators from $\IC_\HC$ to each segment, $(0.E_j]$, which lies entirely within $\ZC_\MC$, implies Theorem~\ref{thm3}, which is what we set out to prove.

\subsection{Locally discriminating the rotated domino states}
The seminal result of \cite{Bennett9} showed that a set of states could be ``non-local" even if each of those states is a tensor product, and thus not entangled. They used a fairly involved (both technically and conceptually) argument to show that the set of nine states on $\HC=\HC_1\otimes\HC_2$ known as the domino states, with each $\HC_j$ of dimension $3$, cannot be perfectly discriminated in $\locc$. The authors in \cite{ChildsLeung} then used a somewhat simplified (but perhaps not quite `simple') argument to prove the same conclusion for a generalization, the rotated domino states, given by
\begin{align}\label{eqn205}
	\ket{\Psi_1}&=\ket{1}\otimes\ket{1}\notag\\
	\ket{\Psi_2}&=\ket{0}\otimes(\cos{\theta_1}\ket{0}+\sin{\theta_1}\ket{1})\notag\\
	\ket{\Psi_3}&=\ket{0}\otimes(\sin{\theta_1}\ket{0}-\cos{\theta_1}\ket{1})\notag\\
	\ket{\Psi_4}&=(\cos{\theta_2}\ket{0}+\sin{\theta_2}\ket{1})\otimes\ket{2}\notag\\
	\ket{\Psi_5}&=(\sin{\theta_2}\ket{0}-\cos{\theta_2}\ket{1})\otimes\ket{2}\notag\\
	\ket{\Psi_6}&=\ket{2}\otimes(\cos{\theta_3}\ket{1}+\sin{\theta_3}\ket{2})\notag\\
	\ket{\Psi_7}&=\ket{2}\otimes(\sin{\theta_3}\ket{1}-\cos{\theta_3}\ket{2})\notag\\
	\ket{\Psi_8}&=(\cos{\theta_4}\ket{1}+\sin{\theta_4}\ket{2})\otimes\ket{0}\notag\\
	\ket{\Psi_9}&=(\sin{\theta_4}\ket{1}-\cos{\theta_4}\ket{2})\otimes\ket{0}
\end{align}
with $0<\theta_j\le\pi/4$ for all $j$ (the original set of \cite{Bennett9} is recovered by setting $\theta_j=\pi/4$ for all $j$). We now give a very simple argument that these states in \myeq{eqn205} cannot be perfectly discriminated within $\locc$. Since these states are a complete, orthogonal basis of $\HC$, the only successful measurement is one consisting of projectors onto each of these nine states, $\MC_{\textrm{dom}}=\{[\Psi_j]\}$. According to Theorem~\ref{thm2}, there must be a continuous, monotonic path of product operators from $I_\HC$ to each outcome of this measurement, and lying in the corresponding zonotope, which we denote as $\ZC_{\textrm{dom}}$. We will show that $(0,I_\HC]$ is an isolated line segment within the intersection of all product operators with $\ZC_{\textrm{dom}}$, which by Corollary~\ref{cor1}, proves that $\MC_\textrm{dom}\not\in\locc$.
\begin{thm5}
	$\MC_\textrm{dom}\not\in\locc$.
\end{thm5}
\proof Consider $z\in \ZC_{\textrm{dom}}$,
\begin{align}\label{eqn206}
	z=\sum_{j=1}^9c_j[\Psi_j],
\end{align}
with $c_j\ge0$ for all $j$. When $z$ is a product operator of the form $\AC\otimes\BC$, the nine $3$-by-$3$ blocks in $z$ must all be proportional to each other. Assuming $z$ is a product operator then, we can write
\begin{align}\label{eqn207}
	\AC_{00}\BC&=\begin{bmatrix}
		c_2\cos^2{\theta_1}+c_3\sin^2{\theta_1}&\frac{1}{2}(c_2-c_3)\sin{2\theta_1}&0\\
		\frac{1}{2}(c_2-c_3)\sin{2\theta_1}&c_2\sin^2{\theta_1}+c_3\cos^2{\theta_1}&0\\
		0&0&c_4\cos^2{\theta_2}+c_5\sin^2{\theta_2}
	\end{bmatrix}
\end{align}
\begin{align}\label{eqn208}
	\AC_{01}\BC&=\begin{bmatrix}
		0&0&0\\
		0&0&0\\
		0&0&\frac{1}{2}(c_4-c_5)\sin{2\theta_2}
	\end{bmatrix}
\end{align}
\begin{align}\label{eqn209}
	\AC_{11}\BC&=\begin{bmatrix}
		c_8\cos^2{\theta_4}+c_9\sin^2{\theta_4}&0&0\\
		0&c_1&0\\
		0&0&c_4\sin^2{\theta_2}+c_5\cos^2{\theta_2}
	\end{bmatrix}
\end{align}
\begin{align}\label{eqn210}
	\AC_{12}\BC&=\begin{bmatrix}
		\frac{1}{2}(c_8-c_9)\sin{2\theta_4}&0&0\\
		0&0&0\\
		0&0&0
	\end{bmatrix}
\end{align}
\begin{align}\label{eqn211}
	\AC_{22}\BC&=\begin{bmatrix}
	c_8\sin^2{\theta_4}+c_9\cos^2{\theta_4}&0&0\\
	0&c_6\cos^2{\theta_3}+c_7\sin^2{\theta_3}&\frac{1}{2}(c_6-c_7)\sin{2\theta_3}\\
	0&\frac{1}{2}(c_6-c_7)\sin{2\theta_3}&c_6\sin^2{\theta_3}+c_7\cos^2{\theta_3}
\end{bmatrix}
\end{align}
One can show that all product operators in $\ZC_{\textrm{dom}}$ are either proportional to $I_\HC$ or have rank no greater than $2$. It is easier, however, to show that all product operators in $\ZC_{\textrm{dom}}$ that are not proportional to $I_\HC$ cannot have full rank equal to $9$ so therefore cannot approach the line segment $(0,I_\HC]$, and this is what we will now do.

If $z=\AC\otimes\BC$ is full rank, then $\AC$ and $\BC$ each have full rank equal to $3$. Notice that Eqs.~\eqref{eqn208} and \eqref{eqn210} preclude $\BC$ having full rank, unless $\AC_{01}=0=\AC_{12}$, so $\AC$ is diagonal, implying that $c_4=c_5$ and $c_8=c_9$. Since $\AC$ is diagonal and full rank, $\AC_{11}\ne0$, so from \myeq{eqn209}, $\BC$ is diagonal, and then from Eqs.~(\ref{eqn207}) and (\ref{eqn211}) we have that $c_2=c_3$ and $c_6=c_7$. Along with the fact that $\AC_{jj}\ne0$ for all $j$, these conditions lead to
\begin{align}\label{eqn212}
	\BC=\frac{1}{\AC_{00}}\begin{bmatrix}
		c_2&0&0\\
		0&c_2&0\\
		0&0&c_4
	\end{bmatrix}
		=\frac{1}{\AC_{11}}\begin{bmatrix}
		c_8&0&0\\
		0&c_1&0\\
		0&0&c_4
	\end{bmatrix}
		=\frac{1}{\AC_{22}}\begin{bmatrix}
			c_8&0&0\\
			0&c_6&0\\
			0&0&c_6
	\end{bmatrix}
\end{align}
From the first expression for $\BC$ we see that $\BC_{00}=\BC_{11}$ and from the third expression, $\BC_{11}=\BC_{22}$, so that $\BC\propto I_B$. Considering these expressions for $\BC_{22}$, we have $c_4/\AC_{00}=c_4/\AC_{11}$ or $\AC_{00}=\AC_{11}$. Similarly considering the expressions for $\BC_{00}$, we find that $\AC_{11}=\AC_{22}$, and therefore, $\AC\propto I_A$, which implies that the only product operators in $\ZC_{\textrm{dom}}$ of rank equal to $9$ are all proportional to $I_\HC$. This means that no non-zero product operators are close to $(0,I_\HC]$ and by Corollary~\ref{cor1}, this completes the proof.\endproof

\subsection{The Unextendible Product Basis known as Tiles}
If, from the domino set of states---those given in \myeq{eqn205} with $\theta_j=\pi/4$ for all $j$---one omits the states $\ket{\Psi_1}$, $\ket{\Psi_2}$, $\ket{\Psi_4}$, $\ket{\Psi_6}$ and $\ket{\Psi_8}$, and adds one extra state given as $\ket{\Psi_{51}}$ in \myeq{eqn301} below, one obtains a set of states known as Tiles. This is an Unextendible Product Basis (UPB), a set of product states for which there is no other product state orthogonal to each of the states in the original set. UPBs have found important applications in quantum information theory \cite{IBM_PRL,IBM_CMP}. As shown in \cite{FuLeungMancinska}, any measurement, $\MC_{\textrm{Tiles}}$, perfectly discriminating the Tiles UPB cannot be implemented within $\locc$. We provide an alternative proof here, as illustration of Theorem~\ref{thm2} (an argument very similar to that we give here can also be used to prove that no UPB on a $3\times3$ system can be perfectly discriminated using $\locc$, a result also obtained in \cite{FuLeungMancinska}). This proof is more difficult than that given in the preceding subsection, since here we are not working with a complete basis of $\HC$, so we must consider a range of possible measurements.

The states of the Tiles UPB are
\begin{align}\label{eqn301}
	\ket{\Psi_{11}}&=\frac{1}{\sqrt{2}}\ket{0}\otimes(\ket{0}-\ket{1})\notag\\
	\ket{\Psi_{21}}&=\frac{1}{\sqrt{2}}(\ket{0}-\ket{1})\otimes\ket{2}\notag\\
	\ket{\Psi_{31}}&=\frac{1}{\sqrt{2}}\ket{2}\otimes(\ket{1}-\ket{2})\\
	\ket{\Psi_{41}}&=\frac{1}{\sqrt{2}}(\ket{1}-\ket{2})\otimes\ket{0}\notag\\
	\ket{\Psi_{51}}&=\frac{1}{3}(\ket{0}+\ket{1}+\ket{2})\otimes(\ket{0}+\ket{1}+\ket{2})\notag
\end{align}
Let us characterize all measurements that perfectly discriminate this set. Each outcome of any such measurement must be orthogonal to all but one of the states. For $\locc$, those measurement outcomes must also be product operators. Consider the local parts of these states (for example, $\ket{0}$ is the first party's local part of $\ket{\Psi_{11}}$ and the second party's local part of $\ket{\Psi_{41}}$). Note that, for either party, the local parts of any three of these states span their local Hilbert space, $\HC_j$. Therefore, no state on $\HC_j$ is orthogonal (on that side) to more than two of the $\ket{\Psi_{i1}}$. This means that any product state orthogonal to four of the states of \myeq{eqn301} must be orthogonal on the $\HC_1$ side to two of the states, and to the other two states on the $\HC_2$ side. For each of the states in Tiles, we can use this observation to identify by inspection six states orthogonal to the other four states in Tiles. Defining
\begin{align}\label{eqn302}
	[i\pm j]&=(\ket{i}\pm\ket{j})(\bra{i}\pm\bra{j})/2\notag\\
	[i\pm j\pm k]&=(\ket{i}\pm\ket{j}\pm\ket{k})\bra{i}\pm\bra{j}\pm\bra{k})/3\notag\\
	[\phi_0]&=(2\ket{0}-\ket{1}-\ket{2})(2\bra{0}-\bra{1}-\bra{2})/6\\
	[\phi_2]&=(\ket{0}+\ket{1}-2\ket{2})(\bra{0}+\bra{1}-2\bra{2})/6\notag
\end{align}
the projectors onto the six states orthogonal to all but $\ket{\Psi_{i1}}$ of the states in Tiles, for each $i$, are
\begin{align}\label{eqn303}
	E_{11}&=[0]\otimes[0-1]&E_{21}&=[0-1]\otimes[2]\notag\\
	E_{12}&=[0+1]\otimes[1-2]&E_{22}&=[1-2]\otimes[1+2]\notag\\
	E_{13}&=[0-1]\otimes[1]&E_{23}&=[1]\otimes[1-2]\notag\\
	E_{14}&=[\phi_0]\otimes[0]&E_{24}&=[0]\otimes[\phi_2]\notag\\
	E_{15}&=[\phi_2]\otimes[1+2]&E_{25}&=[1+2]\otimes[\phi_0]\notag\\
	E_{16}&=[0+1+2]\otimes[\phi_0]&E_{26}&=[\phi_0]\otimes[0+1+2]\notag\\\notag\\
	E_{31}&=[2]\otimes[1-2]&E_{41}&=[1-2]\otimes[0]\notag\\
	E_{32}&=[1+2]\otimes[0-1]&E_{42}&=[0-1]\otimes[0+1]\notag\\
	E_{33}&=[1-2]\otimes[1]&E_{43}&=[1]\otimes[0-1]\notag\\
	E_{34}&=[\phi_2]\otimes[2]&E_{44}&=[2]\otimes[\phi_0]\notag\\
	E_{35}&=[\phi_0]\otimes[0+1]&E_{45}&=[0+1]\otimes[\phi_2]\notag\\
	E_{36}&=[0+1+2]\otimes[\phi_2]&E_{46}&=[\phi_2]\otimes[0+1+2]\notag\\\notag\\
	E_{51}&=[0+1+2]\otimes[0+1+2]\notag\\
	E_{52}&=[0]\otimes[0+1]\notag\\
	E_{53}&=[0+1]\otimes[2]\notag\\
	E_{54}&=[2]\otimes[1+2]\notag\\
	E_{55}&=[1+2]\otimes[0]\notag\\
	E_{56}&=[1]\otimes[1]
\end{align}

Any outcome of $\MC_{\textrm{Tiles}}$ identifying $\ket{\Psi_{i1}}$ must be a linear combination of the $E_{ij}$ for fixed $i$ and $j=1,6$. Noting that the local operators on one (or the other) side of all the $E_{ij}$, for fixed $i$, are linearly independent, the only such linear combinations that are product operators are the individual $E_{ij}$, themselves. Therefore, if $\MC_{\textrm{Tiles}}\in\locc$, each of its outcomes must be proportional to one member in a subset of these (rank-$1$) operators $E_{ij}$. This still leaves us with a range of possible measurements to consider, but our task is simplified by the observation that for any one of these possible measurements, its corresponding zonotope lies within the zonotope defined by this entire set of $30$ projection operators. We will call the latter zonotope $\ZC_{\textrm{Tiles}}$, and then we can prove $\MC_{\textrm{Tiles}}\not\in\locc$ by showing there is no continuous, monotonic path of product operators, lying entirely within $\ZC_{\textrm{Tiles}}$, from $I_\HC$ to any one of these outcomes. We will do this with reference to Corollary~\ref{cor1}, by showing that $(0,I_\HC]$ is isolated within the intersection of $\ZC_{\textrm{Tiles}}$ and the set of product operators.
\begin{thm6}
	The unextendible product basis known as Tiles cannot be perfectly discriminated within  $\locc$.
\end{thm6}
\proof 
Consider an arbitrary point in $\ZC_{\textrm{Tiles}}$,
\begin{align}\label{eqn304}
	R=\sum_{jk}c_{jk}E_{jk},
\end{align}
with $c_{jk}\ge0$ for all $j,k$. We seek conditions under which $R=\AC\otimes\BC$ is a product operator. Note that each of the local projectors appearing in \myeq{eqn303} can be written as a linear combination of the six linearly independent projectors, $[0],[1],[2],[0+1],[1+2]$, and $[0+1+2]$ on either party. (For example, $[0-1]=[0]+[1]-[0+1]$, $[1-2]=[1]+[2]-[1+2]$, $[\phi_0]=[0]+[1+2]-[0+1+2]$, and $[\phi_2]=[2]+[0+1]-[0+1+2]$.) Rewriting the $E_{jk}$ in terms of these six linearly independent local operators on each side, one finds that the following six product operators do not appear anywhere in this set of $30$ operators: 
\begin{align}\label{eqn305}
	&[0]\otimes[1+2]&&[1]\otimes[0+1+2]\notag\\
	&[2]\otimes[0+1]&&[0+1]\otimes[0]\notag\\
	&[1+2]\otimes[2]&&[0+1+2]\otimes[1].
\end{align}
If we rewrite $R$ in the same way, none of these six product operators will appear. We can expand $R$ in terms of the six linearly independent projectors on the $A$-side as
\begin{align}\label{eqn306}
	R=[0]\otimes\BC_{0}+[1]\otimes\BC_{1}+[2]\otimes\BC_{2}+[0+1]\otimes\BC_{3}+[1+2]\otimes\BC_{4}+[0+1+2]\otimes\BC_{5},
\end{align}
where $\BC_{0}$ has no term with $[1+2]$ appearing in it, $\BC_{1}$ has no $[0+1+2]$, etc. Now, in order for $R$ to be a product operator, the $\BC_{i}$ must all be proportional to each other. This means that either $\BC_{0}=0$ or none of the $\BC_i$ contain a term with $[1+2]$; either $\BC_1=0$ or none of the $\BC_i$ contain a term with $[0+1+2]$; and so on. In other words, either $[0]$ appears nowhere on the $A$ side, or $[1+2]$ appears nowhere on the $B$ side; either $[1]$ appears nowhere on the $A$ side, or $[0+1+2]$ appears nowhere on the $B$ side; etc. That is, for each of the six operators appearing in \myeq{eqn305}, either the $A$-part is absent from $\AC$ or the $B$ part is absent from $\BC$.

Next, consider what these observations tell us about the presence of product operators in $\ZC_{\textrm{Tiles}}$ that are close to $I_\HC$. First, note that any such operator must have full rank equal to $9$, so that $\textrm{rank}(\AC)=\textrm{rank}(\BC)=3$. This means neither $\AC$ nor $\BC$ can be missing more than three of the local operators mentioned in the preceding paragraph. Since together, the two must be missing a total of six of these operators, we reach the conclusion that $\AC$ and $\BC$ must each be missing three of them. Considering all possible trios from the six local operators, $[0],[1],[2],[0+1],[1+2],[0+1+2]$, it is easy to see that the only trio that allows $\AC$ ($\BC$) to be close to $I_A$ ($I_B$) is $[0],[1],[2]$. For example, the trio $[0],[1],[0+1]$ gives $\bra{2}\AC\ket{2}=0$ so is not close to $I_A$; the trio 
$[0],[1],[1+2]$ has $\bra{1}\AC\ket{2}=\bra{2}\AC\ket{2}$ so is not close to $I_A$; and so on. Therefore, $\AC$, $\BC$ must be diagonal in the standard basis, $[0],[1],[2]$. This condition imposes strong constraints on the coefficients $\{c_{jk}\}$ in \myeq{eqn304}, which reduce the expression for $R$ to the required diagonal form, with (nonzero) entries $[x~~x~~y~~w~~t~~y~~w~~z~~z]$. The first set of three entries is $\AC_{00}\BC$, so $x\propto\BC_{00}=\BC_{11}$, and the last set of three entries is $\AC_{22}\BC$, so $z\propto\BC_{22}=\BC_{11}$. That is, $\BC$ must be proportional to $I_B$. This tells us that $x=y=w=t=z$, from which we see that $R=xI_\HC$. Therefore, the only product operators in $\ZC_{\textrm{Tiles}}$ close to $(0,I_\HC]$ are proportional to $I_\HC$, itself, and this completes the proof. \endproof

\subsection{The Unextendible Product Basis known as Shifts}
We next turn our attention to the Shifts UPB, which consists of the set of four three-qubit states,
\begin{align}\label{eqn401}
	\ket{\Psi_{11}}&=\ket{000}\notag\\
	\ket{\Psi_{21}}&=\ket{+1-}\notag\\
	\ket{\Psi_{31}}&=\ket{1-+}\notag\\
	\ket{\Psi_{41}}&=\ket{-+1},
\end{align}
where
\begin{align}\label{eqn402}
	\ket{\pm}=\left(\ket{0}\pm\ket{1}\right)/\sqrt{2}.
\end{align}
This UPB can be analyzed using the exact same approach as was used in the preceding subsection for Tiles. The local parts of any two of these states span their local Hilbert space, $\HC_j$. Therefore, no state on $\HC_j$ is orthogonal (on that side) to more than one of the $\ket{\Psi_{i1}}$. This means that any product state orthogonal to three of the states of \myeq{eqn401} must be orthogonal to one of those three states on $\HC_1$, one on $\HC_2$, and the third on $\HC_3$. For each of the states in Shifts, we can use this observation to identify by inspection six states orthogonal to the other three states in Shifts. The projectors onto the six states orthogonal to all but $\ket{\Psi_{i1}}$ of the states in Shifts, for each $i$, are
\begin{align}\label{eqn403}
	E_{11}&=[000]&E_{21}&=[+1-]\notag\\
	E_{12}&=[-+0]&E_{22}&=[1+0]\notag\\
	E_{13}&=[---]&E_{23}&=[1--]\notag\\
	E_{14}&=[0-+]&E_{24}&=[0-1]\notag\\
	E_{15}&=[+++]&E_{25}&=[010]\notag\\
	E_{16}&=[+0-]&E_{26}&=[++1]\notag\\\notag\\
	E_{31}&=[1-+]&E_{41}&=[-+1]\notag\\
	E_{32}&=[100]&E_{42}&=[10-]\notag\\
	E_{33}&=[-10]&E_{43}&=[1++]\notag\\
	E_{34}&=[--1]&E_{44}&=[-1-]\notag\\
	E_{35}&=[+1+]&E_{45}&=[001]\notag\\
	E_{36}&=[+01]&E_{46}&=[01+]
\end{align}

Any outcome of a measurement that perfectly discriminates Shifts, which identifies $\ket{\Psi_{i1}}$, must be a linear combination of the $E_{ij}$ for fixed $i$ and $j=1,6$. Once again, while we have a range of possible measurements to consider, our task is simplified by the fact that for any one of these possible measurements, its corresponding zonotope lies within the zonotope defined by this entire set of $24$ projection operators given in \myeq{eqn403}. Denoting the latter zonotope as $\ZC_{\textrm{Shifts}}$, we can prove Shifts is not in $\locc$ by showing there is no continuous, monotonic path of product operators, lying entirely within $\ZC_{\textrm{Shifts}}$, from $I_\HC$ to any one of these outcomes. Once again, we use Corollary~\ref{cor1}, and show that $(0,I_\HC]$ is isolated within the intersection of $\ZC_{\textrm{Shifts}}$ and the set of product operators.
\begin{thm7}
	The unextendible product basis known as Shifts cannot be perfectly discriminated within  $\locc$.
\end{thm7}
\proof 
Consider an arbitrary point in $\ZC_{\textrm{Shifts}}$,
\begin{align}\label{eqn404}
	R=\sum_{jk}c_{jk}E_{jk}.
\end{align}
We seek conditions under which $R=\AC\otimes\BC\otimes\CC$ is a product operator. Note that each of the local projectors appearing in \myeq{eqn403} can be written as a linear combination of the three linearly independent projectors, $[0],[+]$, and $[-]$.\footnote{We have chosen to eliminate $[1]$ here, because $[111]$ is not one of the $E_{jk}$. If instead we had eliminated $[0]$, for example, then since $E_{11}=[000]$, all possible combinations of $[1],[+],[-]$, such as $[111],[1+-],[+0-]$, etc., would appear in $R$, and the approach of the preceding section would not work. Due to the presence of $E_{13}$ and $E_{15}$, the same conclusion would hold for eliminating $[-]$ or $[+]$.} Rewriting the $E_{jk}$ in terms of these three linearly independent local operators, one finds that the following three product operators do not appear anywhere in this set of $24$ operators: 
\begin{align}\label{eqn405}
	[0+-]&&[+-0]&&[-0+].
\end{align}
If we rewrite $R$ in the same way, none of these three product operators will appear. We can expand $R$ as
\begin{align}\label{eqn406}
	R=[0]\otimes\BC_{0}\otimes\CC_{0}+[+]\otimes\BC_{+}\otimes\CC_{+}+[-]\otimes\BC_{-}\otimes\CC_{-},
\end{align}
where $\BC_{0}\otimes\CC_{0}$ has no term with $[+-]$ appearing in it, $\BC_{+}\otimes\CC_{+}$ has no $[-0]$, and $\BC_{-}\otimes\CC_{-}$ has no $[0+]$. Now, in order for $R$ to be a product operator, the $\BC_{i}\otimes\CC_{i}$ must all be proportional to each other, which also means that all the $\BC_i$ are proportional to each other and all the $\CC_i$ are proportional to each other. This means that either $\BC_{0}\otimes\CC_{0}=0$ or none of the $\BC_i\otimes\CC_{i}$ contain a term with $[+-]$, in turn implying none of the $\BC_i$ has $[+]$ and/or none of the $\CC_{i}$ has $[-]$. Similarly from the $[+]$ on the $A$ side, either $\BC_+\otimes\CC_+=0$ or none of the $\BC_i$ contain a term with $[-]$ and/or none of the $\CC_i$ have $[0]$; and from the $[-]$ on the $A$ side, either $\BC_-\otimes\CC_-=0$ or none of the $\BC_i$ contain a term with $[0]$ and/or none of the $\CC_{i}$ have $[+]$. In other words, for each of the three operators appearing in \myeq{eqn405}, the $A$-part is absent from $\AC$ and/or the $B$ part is absent from $\BC$ and/or the $C$ part is absent from $\CC$.

Next, consider what these observations tell us about the presence of product operators in $\ZC_{\textrm{Shifts}}$ that are close to $I_\HC$. First, note that any such operator must have full rank equal to $8$, so that $\textrm{rank}(\AC)=\textrm{rank}(\BC)=\textrm{rank}(\CC)=2$. This means each of $\AC$, $\BC$, and $\CC$ must have contributions from at least two of the operators $[0],[+]$, and $[-]$, implying none can be missing more than one of these local operators. Now, in order to exclude all three of the operators in \myeq{eqn405}, the three parties combined must be missing a total of at least three of those local operators. Hence, we can conclude that $\AC$, $\BC$, and $\CC$ must each be missing exactly one of them, so each must be a linear combination of exactly two of them (with non-vanishing coefficients). Considering all possible pairs from the three local operators, $[0],[+],[-]$, it is easy to see that the only pair that allows $\AC$ (or $\BC$ or $\CC$) to be close to $I_A$ (or $I_B$ or $I_C$) is $[+],[-]$. Therefore, $\AC$, $\BC$, and $\CC$ must each be diagonal in the $[+],[-]$ basis, which constrains the coefficients $\{c_{jk}\}$ in \myeq{eqn404} and reduces the expression for $R$ to the required diagonal form,
\begin{align}\label{eqn407}
	\AC\otimes\BC\otimes\CC=[+]\otimes\left(t[++]+x[+-]+y[-+]+x[--]\right)+[-]\otimes\left(z[++]+z[+-]+y[-+]+w[--]\right)
\end{align}
The two expressions in parentheses must be proportional to each other in order for this to be a product operator. From the $[-+]$ terms, we see that the two expressions must actually be equal. Therefore, $x=z=t=w$, which reduces this to
\begin{align}\label{eqn408}
	\AC\otimes\BC\otimes\CC=I_A\otimes\left(x[+]\otimes I_C+[-]\otimes\left(y[+]+x[-]\right)\right).
\end{align}
This is not a product operator unless $x=y$, leaving us with $\AC\otimes\BC\otimes\CC=xI_\HC$. That is, the only product operators in $\ZC_{\textrm{Shifts}}$ that are close to $(0,I_\HC]$ are proportional to $I_\HC$, itself, and this completes the proof. \endproof

\subsection{The Class of Unextendible Product Bases known as GenTiles$2$}
\begin{figure}
	\includegraphics[scale=1]{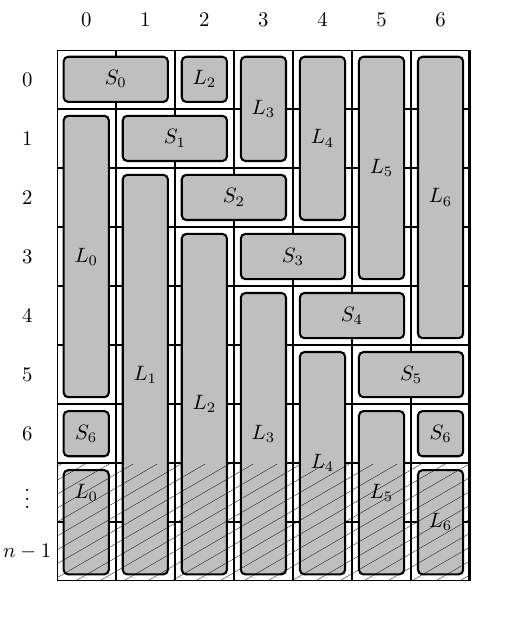}\caption{\label{fig103}GenTiles$2$ on a $7\times n$ system. The shaded area at the bottom of the figure represents extension of those tiles to an arbitrary length $n-2$.}
\end{figure}

There are two generalizations of the Tiles UPB, GenTiles$1$ and GenTiles$2$, each of which are infinite classes of UPB's on bipartite systems that have a tiling representation reminiscent of Tiles. We first discuss GenTiles$2$, leaving GenTiles$1$ to the next subsection. GenTiles$2$ is a UPB on an $m\times n$ Hilbert space $\HC$, which for all $n>3$, $m\ge3$, and $n\ge m$, consists of states
\begin{align}\label{eqn3001}
	\ket{S_{j}}&=\frac{1}{\sqrt{2}}\left(\ket{j}-\ket{j+1\mmod{m}}\right)\otimes\ket{j}\notag\\
	\ket{L_{jk}}&=\ket{j}\otimes\frac{1}{\sqrt{n-2}}\left(\sum_{i=0}^{m-3}\omega^{ik}\ket{i+j+1\mmod{m}}+\sum_{i=m-2}^{n-3}\omega^{ik}\ket{i+2}\right),\\
	\ket{F}&=\frac{1}{\sqrt{nm}}\sum_{i=0}^{m-1}\sum_{j=0}^{n-1}\ket{i}\otimes\ket{j},\notag
\end{align}
with $0\le j\le m-1$, $1\le k\le n-3$, and $\omega=e^{2\pi i/(n-2)}$. We will also use the states $\ket{T_{j}}=\left(\ket{j}+\ket{j+1\mmod{m}}\right)\otimes\ket{j}/\sqrt{2}$ and $\ket{L_{j0}}$ in our arguments, even though they are not a part of the UPB. The states $\ket{L_{jk}}$ with fixed $j$ share the same support and all lie within one of the long tiles of length $n-2$ (and labeled as $L_0$ to $L_6$ for the $m=7$ case) in Figure~\ref{fig103}. The short tiles representing the $\ket{S_j}$, labeled $S_0$ to $S_6$ in the figure, are each of length two, and the state $\ket{F}$, which covers the entire figure, is not shown.

Here we prove the theorem,
\begin{thm9}\label{thm7}
	The unextendible product basis known as GenTiles$2$ cannot be perfectly discriminated within $\locc$.
\end{thm9}
\noindent As usual, the first step is to identify possible separable measurements that accomplish perfect discrimination. Toward this end, we prove the lemma,
\begin{lem11}\label{lem11}
		The only product states in $\HC$ that are orthogonal to all but one of the states in GenTiles$2$ are $\ket{F}$ and the sets of states $\{\ket{S_{j}}\}$, $\{\ket{T_j}\}$, $\{\ket{L_{jk}}\}$, $\sqrt{2}\ket{L_{j0}}-\sqrt{n-2}\left(\ket{T_{j-1}}-\ket{S_{j-1}}\right)$, and $\sqrt{2}\ket{L_{j0}}-\sqrt{n-2}\left(\ket{T_{j}}+\ket{S_{j}}\right)$, for $j=0,\cdots,m-1$ and $k=0,\cdots,n-3$, unless $n=4$. When $n=4$, states $\left(\ket{0}+\ket{1}+\ket{2}\right)\otimes\left(\sum_{i\ne j}^{3}\ket{i}-3\ket{j}\right)$ are also allowed when $m=3$; $\ket{L_{j0}}+\ket{L_{j1}}-\ket{T_{j+1}}$ are also allowed when $m=3,4$; and $\ket{L_{j0}}-\ket{L_{j1}}-\ket{T_{j+2}}$ are also allowed when $m=4$. [All indices are to be understood as $\mmod{m}$ and $j=0,\cdots,m-1$.]
\end{lem11}
\noindent The proof is given in Appendix~\ref{AppB1}. Given this result, we then prove in Appendix~\ref{AppB2}, that
\begin{lem18}\label{lem18}
	The only complete separable measurement that perfectly discriminates GenTiles$2$ consists of projectors onto the states, $\ket{S_j},\ket{T_j}$, and $\ket{L_{jk}}$, for $j=0,\cdots,m-1$ and $k=0,\cdots,n-3$.
\end{lem18}

We are now in a position to prove Theorem~\ref{thm7}.
\proof From Lemma~\ref{lem18}, we have that the only separable measurement that perfectly discriminates GenTiles$2$ is $\MC_{GT2}=\{[S_j],[T_j],[L_{jk}]\}$ for $j=0,\cdots,m-1$ and $k=0,\cdots,n-3$, and its associated zonotope will be denoted $\ZC_{GT2}$. The proof of this theorem will use Corollary~\ref{cor1} and closely follows the approach of the preceding subsections. %As such, define $E_i=[S_i]$, $E_{m+i}=[T_i]$, and $E_{(k+2)m+i}=[L_{ik}]$. Further,
Defining $\ket{L_{jk}}=\ket{j}\otimes\ket{l_{jk}}$, so that 
\begin{align}\label{eqn3000}
	\ket{l_{jk}}=\frac{1}{\sqrt{n-2}}\left(\sum_{i=0}^{m-3}\omega^{ik}\ket{i+j+1\mmod{m}}+\sum_{i=m-2}^{n-3}\omega^{ik}\ket{i+2}\right),
\end{align}
\noindent and $[j_\pm]=\left(\ket{j}\pm\ket{j+1\mmod{m}}\right)\left(\bra{j}\pm\bra{j+1\mmod{m}}\right)/2$, any operator $R\in\ZC_{GT2}$ may be written
\begin{align}\label{eqn3002}
	R&=\sum_{j=0}^{m-1}\left(a_j[S_j]+b_j[T_j]+\sum_{k=0}^{n-3}c_{jk}[L_{jk}]\right)\notag\\
	&=\sum_{j=0}^{m-1}\left([j]\otimes\BC_j+[j_-]\otimes\BC_{m+j}\right),
\end{align}
with $c_i\ge0$ for all $i$. We have chosen $\{[j],[j_-]\}_{j=0}^{m-1}$ because it constitutes a linearly independent set, which means that the set of operators, $\{\BC_i\}_{i=0}^{2m-1}$, must all be proportional to each other in order that $R$ is a product operator, as we require for the application of Corollary~\ref{cor1}. Note also that $[j_+]=[j]+[j+1\mmod{m}]-[j_-]$.

Expanding, we have
\begin{align}\label{eqn3003}
	R&=\sum_{j=0}^{m-1}\left[a_j[j_-]\otimes[j]+b_j\left([j]+[j+1\mmod{m}]-[j_-]\right)\otimes[j]+[j]\otimes\sum_kc_{jk}[l_{jk}]\right]\notag\\
		&=\sum_{j=0}^{m-1}\left[[j]\otimes\left(b_j[j]+\sum_kc_{jk}[l_{jk}]\right)+b_j[j+1\mmod{m}]\otimes[j]+[j_-]\otimes\left(a_j[j]-b_j[j]\right)\right]\notag\\
		&=\sum_{j=0}^{m-1}\left[[j]\otimes\left(b_j[j]+b_{j-1}[j-1\mmod{m}]+\sum_kc_{jk}[l_{jk}]\right)+\left(a_j-b_j\right)[j_-]\otimes[j]\right]
\end{align}
From this, we identify $\BC_j=b_j[j]+b_{j-1}[j-1\mmod{m}]+\sum_kc_{jk}[l_{jk}]$ and $\BC_{m+j}=\left(a_j-b_j\right)[j]$ for $j=0,\cdots,m-1$. Since all the $\BC_i$ must be proportional to each other, they must all have rank equal to $1$, as do all the $\BC_{m+j}$, unless $a_j=b_j$ for all $j$. To use Corollary~\ref{cor1}, we seek all $R\in\ZC_{GT2}$ that are close to $I_\HC$, so that rank($R$)$=mn$ and rank($\BC_j$)$=n$ for all $j$. Thus, $a_j=b_j$, and
\begin{align}\label{eqn3004}
	R&=\sum_{j=0}^{m-1}[j]\otimes\left(b_j[j]+b_{j-1}[j-1\mmod{m}]+\sum_kc_{jk}[l_{jk}]\right).
\end{align}
Noting that $\inpd{j}{l_{jk}}=0=\inpd{j-1\mmod{m}}{l_{jk}}$, compare
\begin{align}\label{eqn3005}
	\BC_0&=b_0[0]+b_{m-1}[m-1]+\sum_kc_{0k}[l_{0k}],\notag\\
	\BC_1&=b_1[1]+b_0[0]+\sum_kc_{1k}[l_{1k}].
\end{align}
These must be proportional to each other, but the coefficients of $[0]$ are equal to $b_0$ in both cases, indicating that in fact, $\BC_0=\BC_1$. By similarly comparing each pair of operators $\BC_j,\BC_{j+1},j=0,\cdots,m-2$ in succession, we easily see that $\BC_j=\BC$, independent of $j$. This indicates that $R=I_{\HC_1}\otimes\BC$ for $R$ a full rank product operator in $\ZC_{GT2}$. Note that $\bra{i}\BC_j\ket{j}=b_j\delta_{ij}$, which tells us that $\BC_j$, and therefore $\BC$, are diagonal in the standard basis. Therefore, $\sum_kc_{jk}[l_{jk}]$ must also be diagonal in that basis.

Expand
\begin{align}\label{eqn3006}
	(n-2)\sum_kc_{jk}[l_{jk}]&=\sum_{i,i^\prime=0}^{m-3}\left(\sum_kc_{jk}\omega^{(i-i^\prime)k}\right)\ket{i+j+1\mmod{m}}\bra{{i^\prime+j+1\mmod{m}}}\notag\\
			&+\sum_{i,i^\prime=m-2}^{n-3}\left(\sum_kc_{jk}\omega^{(i-i^\prime)k}\right)\ket{i+2}\bra{i^\prime+2}\notag\\
			&+\left[\sum_{i=0}^{m-3}\sum_{i^\prime=m-2}^{n-3}\left(\sum_kc_{jk}\omega^{(i-i^\prime)k}\right)\ket{i+j+1\mmod{m}}\bra{i^\prime+2}+h.c.\right].
\end{align}
where $h.c.$ stands for Hermitian conjugate, and the sums over $k$ run from $k=0$ to $k=n-3$. In order that $\BC_j$ is diagonal in the standard basis, \myeq{eqn3006} tells us that $\sum_kc_{jk}\omega^{pk}=0$ for $p=1,\cdots,n-3$. This is only possible if $c_{jk}=c_{j0}$, independent of $k$. Then, \myeq{eqn3006} reduces to
\begin{align}\label{eqn3007}
	\sum_kc_{jk}[l_{jk}]&=c_{j0}\left(\sum_{i=0}^{m-3}[i+j+1\mmod{m}]+\sum_{i,=m-2}^{n-3}[i+2]\right)=c_{j0}\left(I_{\HC_2}-[j-1\mmod{m}]-[j]\right).
\end{align}
This leaves us with
\begin{align}\label{eqn3008}
	\BC_j&=(b_j-c_{j0})[j]+(b_{j-1}-c_{j0})[j-1\mmod{m}]+c_{j0}I_{\HC_2}.
\end{align}
Comparing all the $\BC_j$ and recalling they must be independent of $j$, we see that $b_j=c_{j0}=b_{j-1}$ for all $j$, implying these coefficents are all independent of $j$, as well, and we have that $\BC_j\propto I_{\HC_2}$. Thus, $R\propto I_\HC$ are the only product operators close to $I_\HC$ in $\ZC_{GT2}$ and by Corollary~\ref{cor1}, this completes the proof.\endproof

\subsection{The Class of Unextendible Product Bases known as GenTiles$1$}
\begin{figure}
	\includegraphics[scale=1]{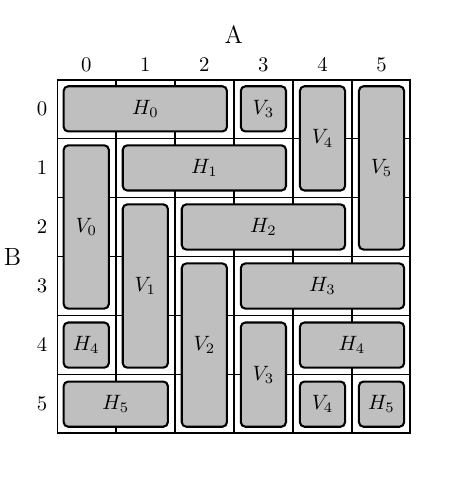}\caption{\label{fig101}GenTiles$1$ on a $6\times6$ system.}
\end{figure}

GenTiles$1$ is an unextendible product basis (UPB) on an $n\times n$ Hilbert space $\HC$, which for all even $n\ge4$ consists of states
\begin{align}\label{eqn1001}
	\ket{V_{km}}&=\frac{1}{\sqrt{n}}\ket{k}\otimes\sum_{j=0}^{\frac{n}{2}-1}\omega^{jm}\ket{j+k+1\mmod{n}},\notag\\
	\ket{H_{km}}&=\frac{1}{\sqrt{n}}\sum_{j=0}^{\frac{n}{2}-1}\omega^{jm}\ket{j+k\mmod{n}}\otimes\ket{k},\\
	\ket{F}&=\frac{1}{n}\sum_{ij=0}^{n-1}\ket{i}\otimes\ket{j},\notag
\end{align}
with $m=1,\cdots,n/2-1$ and $k=0,\cdots,n-1$ and $\omega=e^{4\pi i/n}$. We will also use the states $\ket{V_{k0}}$ and $\ket{H_{k0}}$ in these arguments, even though they are not a part of the UPB. The states $\ket{V_{km}}$ with fixed $k$ share the same support, lying within the $V_k$-tile shown (for the $n=6$ case) in Figure~\ref{fig101}. Similarly the $H_k$-tile shown in that figure contains all the states $\ket{H_{km}}$ for fixed $k$. These tiles are all of length $n/2$. 

Here we prove the theorem,
\begin{thm9}\label{thm9}
		The unextendible product basis known as GenTiles$1$ cannot be perfectly discriminated within $\locc$.
\end{thm9}
\noindent We will see that for each $n$, there is one and only one measurement that accomplishes perfect discrimination of the given set of states. For this complete measurement $\MC_\textrm{GT$1$}$ and the zonotope $\ZC_\textrm{GT$1$}$ that it generates, we find once again that $(0,I_\HC]$ is an isolated line segment within the intersection of $\ZC_\textrm{GT$1$}$ with the set of all product operators acting on $\HC$, which by Corollary~\ref{cor1}, is what we need to prove this theorem. Note that the case of GenTiles$1$ with $n=4$ is identical to GenTiles$2$ on a $4\times4$ system, so we will here confine the discussion to even $n\ge6$.

We start with the following lemma telling us which product states may be included in one of these measurements.
\begin{lem8}\label{lem8}
	The only product states in $\HC$ that are orthogonal to all but one of the states in GenTiles$1$ are $\ket{F}$ and the sets of states $\{\ket{H_{km}}\}$ and $\{\ket{V_{km}}\}$, for $k=0,\cdots,n-1$ and $m=0,\cdots,n/2-1$ ($n\ge6$).
\end{lem8}
\noindent The next lemma tells us what the unique separable measurement is that serves our present purposes.
\begin{lem6}\label{lem6}
	The sole complete separable measurement, $\MC_\textrm{GT$1$}$, that perfectly distinguishes GenTiles$1$ consists of projectors onto the states $\{\ket{H_{km}}\}$ and $\{\ket{V_{km}}\}$, for $k=0,\cdots,n-1$ and $m=0,\cdots,n/2-1$.
\end{lem6}
\noindent Lemmas~\ref{lem8} and \ref{lem6} may be proven using arguments that are entirely analogous to the proofs of Lemmas~\ref{lem11} and \ref{lem18} for GenTiles$2$, respectively, so we omit the details here.

The proof that $\MC_{GT1}\not\in\locc$ is more challenging than others we've given, mainly due to the difficulty in identifying an appropriate, simple, linearly independent set of operators, as was done for the sets of states considered in preceding subsections. Therefore, we leave the proof to Appendix~\ref{AppA3}, where we show that the only product operators in $\ZC_\textrm{GT$1$}$ that have rank greater than $n/2$ are all proportional to $I_\HC$. Therefore, by Corollary~\ref{cor1}, we are led to the conclusion in Theorem~\ref{thm9}.

\subsection{Minimum error discrimination: The double trine ensemble of Peres and Wootters}
The preceding examples have involved perfect discrimination of quantum states. Here, we consider a different case, that of achieving the minimum possible error in discriminating the double trine ensemble, originally discussed in the seminal paper of Peres and Wootters \cite{PeresWootters}. The double trine ensemble consists of states $\ket{\psi_i}\otimes\ket{\psi_i}$ with $\ket{\psi_i}=U^i\ket{0}$ and $U=\exp(-i\pi\sigma_y/3)$ with $\sigma_y$ the usual Pauli operator. It was shown in \cite{ChitambarHsiehPeresWootters} that for discriminating the double trine ensemble, the minimum error achievable using global operations can only be achieved using $\locc$ if the following orthogonal set of states can be \emph{perfectly} discriminated using $\locc$:
\begin{align}\label{eqn8001}
	\ket{F_1}&=\frac{1}{\sqrt{6}}\left[\left(\sqrt{2}+1\right)\ket{00}-\left(\sqrt{2}-1\right)\ket{11}\right]\notag\\
	\ket{F_2}&=\frac{1}{3}\left[\left(\sqrt{2}-1\right)\ket{00}+\sqrt{3}\left(\ket{01}+\ket{10}\right)+\left(\sqrt{2}+1\right)\ket{11}\right]\notag\\
	\ket{F_3}&=\frac{1}{3}\left[\left(\sqrt{2}-1\right)\ket{00}-\sqrt{3}\left(\ket{01}+\ket{10}\right)+\left(\sqrt{2}+1\right)\ket{11}\right]
\end{align}
\noindent The singlet state, $\left(\ket{01}-\ket{10}\right)/\sqrt{2}$, is orthogonal to each of the $\ket{F_i}$. Consistent with the discussion in \cite{ChitambarHsiehPeresWootters}, we find via a straightforward argument that any (refined) separable measurement perfectly distinguishing the states of \myeq{eqn8001} must consist of rank-$1$ operators proportional to projectors onto the (unnormalized) product states,
\begin{align}\label{eqn8002}
	\ket{\phi_{11}}&=\left(x\ket{0}+\ket{1}\right)\otimes\left(x\ket{0}-\ket{1}\right)\notag\\
	\ket{\phi_{12}}&=\left(x\ket{0}-\ket{1}\right)\otimes\left(x\ket{0}+\ket{1}\right)\notag\\
	\ket{\phi_{21}}&=\left(z\ket{0}+x\ket{1}\right)\otimes\left(y\ket{0}+x\ket{1}\right)\notag\\
	\ket{\phi_{22}}&=\left(y\ket{0}+x\ket{1}\right)\otimes\left(z\ket{0}+x\ket{1}\right)\notag\\
	\ket{\phi_{31}}&=\left(z\ket{0}-x\ket{1}\right)\otimes\left(y\ket{0}-x\ket{1}\right)\notag\\
	\ket{\phi_{32}}&=\left(y\ket{0}-x\ket{1}\right)\otimes\left(z\ket{0}-x\ket{1}\right),
\end{align}
with $x=\sqrt{2}+1$, $y=\sqrt{3}+\sqrt{2}$, and $z=\sqrt{3}-\sqrt{2}$. We note that $I_\HC$ does in fact lie within the zonotope defined by this set of operators, $\ZC_\MC$; the corresponding separable measurement, $\MC$, perfectly distinguishes the states of \myeq{eqn8001}. We will show that there are no product operators in $\ZC_\MC$ that are arbitrarily close to $I_\HC$, which by our Corollary~\ref{cor1} provides an alternative proof that the globally achievable minimum error for discriminating the double trine ensemble cannot be achieved using $\locc$, a result previously obtained in \cite{ChitambarHsiehPeresWootters}.

Each point in $\ZC_\MC$ is of the form $R=\sum_{ij}c_{ij}\phi_{ij}$ with $\phi_{ij}=\ket{\phi_{ij}}\bra{\phi_{ij}}$. We seek $R=\AC\otimes\BC$ and close to $I_\HC$. Writing $\AC_{kl}\BC_{mn}=\bra{km}R\ket{ln}$, $\hat c_1=c_{11}+c_{12},\hat c_2=c_{21}+c_{31},\hat c_3=c_{22}+c_{32},\hat c_4=c_{11}-c_{12},\hat c_5=c_{21}-c_{31},\hat c_6=c_{22}-c_{32}$, we have
\begin{align}\label{eqn8003}
	\AC_{00}\BC_{00}&=x^4\hat c_1+\hat c_2+\hat c_3\notag\\
	\AC_{00}\BC_{11}&=x^2\left(\hat c_1+z^2\hat c_2+y^2\hat c_3\right)\notag\\
	\AC_{01}\BC_{01}&=x^2\left(-\hat c_1+\hat c_2+\hat c_3\right)\notag\\
	\AC_{11}\BC_{00}&=x^2\left(\hat c_1+y^2\hat c_2+z^2\hat c_3\right)\notag\\
	\AC_{11}\BC_{11}&=\hat c_1+x^4\left(\hat c_2+\hat c_3\right)\notag\\
	\AC_{00}\BC_{01}&=-x^3\hat c_4+x\left(z\hat c_5+y\hat c_6\right)\notag\\
	\AC_{01}\BC_{00}&=x^3\hat c_4+x\left(y\hat c_5+z\hat c_6\right)\notag\\
	\AC_{01}\BC_{11}&=x\hat c_4+x^3\left(z\hat c_5+y\hat c_6\right)\notag\\
	\AC_{11}\BC_{01}&=-x\hat c_4+x^3\left(y\hat c_5+z\hat c_6\right).
\end{align}
Note that $\AC_{01}=\AC_{10}$ and $\BC_{01}=\BC_{10}$ and these are real numbers because the $\ket{\phi_{ij}}$ are real. The expressions in \myeq{eqn8003} can be manipulated to find three independent constraints that $R$ is a product operator. Defining $\alpha_0=\AC_{01}/\AC_{00},\alpha_1=\AC_{11}/\AC_{00},\beta_0=\BC_{01}/\BC_{00},\beta_1=\BC_{11}/\BC_{00}$, these constraints are
\begin{align}\label{eqn8004}
	x^2&=\left(1-x^4\right)\alpha_0\beta_0+x^2\alpha_1\beta_1\notag\\
	\beta_1+\alpha_1&=2+4x\alpha_0\beta_0\notag\\
	x^2\left(\beta_0+\alpha_0\right)&=\alpha_0\beta_1+\alpha_1\beta_0.
\end{align}
From these we find that
\begin{align}\label{eqn8005}
	\alpha_0^2&=\frac{x^4+x^2\alpha_1\left[x^2\left(\alpha_1-2\right)-\left(\alpha_1-1\right)^2\right]}{x^6-2x^4+4x^3-x^2+2-\alpha_1\left(4x^5-x^4+1\right)}.
\end{align}

Note that $R=I_\HC$ corresponds to $\alpha_1=1$ and $\alpha_0=0$. We now ask if $R$ as a product operator can be close to $I_\HC$. To see if this is possible, set $\alpha_1=1+\epsilon$ and $\alpha_0^2=\eta>0$, with $|\epsilon|<<1,\eta<<1$. Inserting these into \myeq{eqn8005} and using $x^2-1=2x$, we find
\begin{align}\label{eqn8006}
	\eta=-\frac{2\epsilon^2x^3}{56+40\sqrt{2}}+\OC(\epsilon^3)
\end{align}
which is manifestly negative, a contradiction (we have inserted $x=\sqrt{2}+1$ into the denominator to simplify the expression). Thus, there are no product operators within $\ZC_\MC$ that are close to $I_\HC$ and $\MC\not\in\locc$.

\subsection{Examples involving optimal unambiguous discrimination}\label{subsecG}
In this subsection, we consider another situation different from all those previously studied here, wherein a set of states is to be unambiguously discriminated, which means that failure is allowed, but the parties must know when they have succeeded in conclusively identifying the state, and when they have failed (an inconclusive result) \cite{CheflesBarnett,CheflesLOCC}. We want to know if the global optimum success probability can be achieved using $\locc$. 

The class we consider here was introduced in \cite{myUSD}. Each member of this class provides a set of states to be unambiguously discriminated, along with the unique separable measurement that achieves the global optimal (minimum) failure rate. We showed in \cite{myUSD} that this measurement cannot be implemented by finite-round LOCC. Here, we show that it also cannot be implemented in $\locc$, excluding the possibility that with an infinite number of rounds, one might come arbitrarily close to achieving the optimal failure rate. The set of states to be discriminated, $\{\ket{\Phi_i}\}$, is reciprocal to the set, $S_\Psi=\{\ket{\Psi_j}\}$, defined below. By reciprocal, we mean that
\begin{align}\label{eqn20}
	\inpd{\Psi_j}{\Phi_i}=\delta_{ij}\inpd{\Psi_i}{\Phi_i}
\end{align}
for all $i,j$ in $S_\Psi$, which must be a linearly independent set. This can be any subset of $D=N-1$ members of the states we are about to describe, for any appropriately chosen local dimensions; see below.

Consider any prime number $N\ge5$ and a multipartite system having overall dimension $D=N-1$. The number of parties $P$ can be chosen in any way consistent with the prime factorization of $D$---this choice is generally not unique, but it is unimportant for our present purposes. Let $\HC_\alpha$ be the Hilbert space describing party $\alpha$'s subsystem, and the overall Hilbert space is then $\HC=\HC_1\otimes\HC_2\otimes\ldots\otimes\HC_P$. Define states
\begin{align}\label{eqn21}
	\ket{\Psi_j}=\ket{\psi_j^{(1)}}\otimes\ldots\otimes\ket{\psi_j^{(P)}},~j=1,\ldots,N,
\end{align}
\noindent with
\begin{align}\label{eqn22}
	\ket{\psi_j^{(\alpha)}}=\frac{1}{\sqrt{d_\alpha}}\sum_{m_\alpha=0}^{d_\alpha-1}e^{2\pi \textrm{i}jp_\alpha m_\alpha/N}\ket{m_\alpha},
\end{align}
\noindent where $d_\alpha$ is the dimension of $\HC_\alpha$, with parties ordered such that $d_1\le d_2\le\cdots\le d_P$, and overall dimension $D=d_1d_2\cdots d_P$. 
Here, $p_1=1$ and for $\alpha\ge2$, $p_\alpha=d_1d_2\cdots d_{\alpha-1}$, and $\ket{m_\alpha}$ is the standard basis for party $\alpha$. It was shown in \cite{myExtViolate1} that any proper subset of these states constitutes a linearly independent set,  and that
\begin{align}\label{eqn23}
	I=\frac{D}{N}\sum_{j=1}^N\Psi_j,
\end{align}
where $\Psi_j=\ket{\Psi_j}\bra{\Psi_j}$. Therefore, by choosing $D=N-1$ of these states---omitting state $J$, say---$\MC_\Psi=\{\Psi_j\}$ is a complete separable measurement that achieves the optimal failure rate, with $\Psi_J$ being the outcome indicating failure. In Appendix~\ref{AppC}, we prove
\begin{thm11}\label{thm11}
	For any choice of $J$ and set of local dimensions, $d_\alpha$, consistent with the overall dimension $D=N-1$, where $N$ is any prime number, the set of states $\{\ket{{\Phi_i}}\}_{i\ne J}$ defined by \myeq{eqn20} cannot be optimally unambiguously discriminated within $\locc$.
\end{thm11}

\section{Conclusion}
In summary, we have presented Theorem \ref{thm2} that if measurement ${\cal M}\in\overline{\textrm{LOCC}}$, then there exists a continuous, monotonic path of positive semidefinite product operators from identity operator ${\cal I}_{\cal H}$ to $(0,E_j]$ for each of the outcomes $E_j$ of ${\cal M}$. We have used Theorem~\ref{thm2} to answer a number of long-standing unsolved problems, including several cases of unextendible product bases for which we have shown they cannot be discriminated within $\overline{\textrm{LOCC}}$. We showed that Proposition $1$ of \cite{KKB} follows from our Theorem~\ref{thm2}, and then used the example from Eq.~$(15)$ of \cite{KKB} to demonstrate that Theorem~\ref{thm2} is strictly stronger than their proposition. Note also that Theorem~\ref{thm2} can be applied to the local state discrimination problem even when $\cap_\mu\ker\rho_\mu$ contains non-vanishing product operators, a circumstance for which Proposition $1$ cannot be used. In addition, Proposition $1$ of \cite{KKB} implies the necessary condition for $\overline{\textrm{LOCC}}$ of \cite{ChildsLeung} that their nonlocality constant $\eta$ vanishes. Therefore, Theorem~\ref{thm2} implies, and is strictly stronger than, the latter necessary condition, as well. Another necessary condition---for perfect discrimination of a pair of multipartite quantum states by $\locc$---has been obtained as Theorem~$1$ in \cite{ChitambarHsiehHierarchy}, and it is also possible to derive this condition from our Theorem~\ref{thm2}. We suspect, though have no proof, that our theorem is strictly stronger than that of \cite{ChitambarHsiehHierarchy}, as well. In any case, since their Theorem~$1$ is restricted to discrimination of pairs of states, our result is certainly much more general.\footnote{Indeed, Theorem~\ref{thm2} applies to all measurements and is not restricted to the local discrimination problem.} Thus, we have succeeded in unifying, and going beyond, all of these previous results on $\overline{\textrm{LOCC}}$. Significantly, the geometric nature of our theorem provides a simple, intuitive way to understand $\overline{\textrm{LOCC}}$.

The question can be raised as to whether or not these necessary conditions for $\overline{\textrm{LOCC}}$ may also be sufficient. The demonstration above that our Theorem~\ref{thm2} is strictly stronger than Proposition $1$ of \cite{KKB} shows that neither Proposition $1$, or the condition of \cite{ChildsLeung} that $\eta=0$, is sufficient. We do not believe that Theorem~\ref{thm2} is sufficient, either, though it remains a possibility. If it is not sufficient, then neither is the necessary condition of \cite{ChitambarHsiehHierarchy}. In our proof of Lemma~\ref{lem5}, we only needed to use a single path to the one outcome $E_{\hat\jmath}$ to show that the conditions of Proposition~1 of \cite{KKB} are satisfied. Our Theorem~\ref{thm2} requires there exists such a path to each of the outcomes, which is, in itself, a stronger requirement. We believe, however, that it is likely that a single path to each outcome is not sufficient for $\overline{\textrm{LOCC}}$, though such cases do exist.\footnote{A simple example of an LOCC measurement for which one, and only one, path exists in $\ZC_{\cal M}$ to each of the outcomes is the measurement on two qubits consisting of outcomes $[0]\otimes[0],[0]\otimes[1],[1]\otimes[+],[1]\otimes[-]$, where $\vert{\pm}\rangle=(\vert{0}\rangle\pm\vert{1}\rangle)/\sqrt{2}$. In this case, the only product operators in $\ZC_{\cal M}$ are those that lie on the single, piecewise local paths to each of these outcomes.} Rather, given that LOCC trees generally involve each branch giving rise to an entire series of subsequent branches, it is likely that some kind of ``tree" of paths to the collection of outcomes would be needed in general for $\overline{\textrm{LOCC}}$, which would mean that the single path of Theorem~\ref{thm2} is not sufficient. Of course, it would not be surprising for the existence of a single path to generally imply the existence of many, but we do not know if there are cases where one or more paths to each outcome exist, while at the same time, the measurement is nonetheless not in $\overline{\textrm{LOCC}}$.
 
Finally, we note that the geometric nature of Theorem~\ref{thm2} points toward ways of extending these results. In particular, the fact that the paths of this theorem must lie within the zonotope $\ZC_{\cal M}$ suggests new approaches for obtaining lower bounds on the error necessarily incurred when implementing a measurement ${\cal M}$ by LOCC. Indeed, as a preliminary result in this direction, we have devised a method of finding such a lower bound when ${\cal M}$ is used to discriminate any orthogonal set of pure states in Hilbert space ${\cal H}$. These results, and hopefully extension to more general sets of states and more general tasks, will be discussed elsewhere.

\noindent\textit{Acknowledgments} --- We are grateful to Dan Stahlke for a series of extremely helpful discussions.

\appendix
\section{Proofs for GenTiles$2$}
\subsection{All product states orthogonal to all but one of the states in GenTiles$2$}\label{AppB1}
	Here, we prove Lemma~\ref{lem11} by finding all product states orthogonal to all but one of the states in GenTiles$2$. Reference will be made to Figure~\ref{fig103} of the main text, along with the indexing of the rows and columns shown there. We use the orthonormal basis of $\HC$ consisting of states $\ket{L_{jk}}$, $\ket{S_{j}}$ and $\ket{T_{j}}$, with $j=0,\cdots,m-1$ and $k=0,\cdots,n-3$. Since the states we seek must be product, they are of the form
	\begin{align}\label{eqn6000}
		\ket{\phi}=\ket{x}\ket{y}=\sum_{j=0}^{m-1}\left(\sum_{k=0}^{n-3}c_{jk}\ket{L_{jk}}+a_{j}\ket{S_{j}}+b_j\ket{T_j}\right).
	\end{align}
	For orthogonality to each state $\ket{S_{j}}$ or $\ket{L_{jk}}$ of GenTiles$2$, we require $a_{j}=0$ or $c_{jk}=0$, respectively. Denote the state that we are not requiring $\ket{\phi}$ to be orthogonal to as $\ket{R}$. Then,
	\begin{align}\label{eqn6001}
		\ket{\phi}=\ket{x}\ket{y}=r\ket{R}+\sum_{j=0}^{m-1}\left(c_{j0}\ket{L_{j0}}+b_{j}\ket{T_{j}}\right),
	\end{align}
	and orthogonality to $\ket{F}$ requires $\sum_{j}(\sqrt{n-2}c_{j0}+\sqrt{2}b_{j}=0)$, as well.
	
	Write $\ket{x}=\sum_ix_i\ket{i}$ and $\ket{y}=\sum_jy_j\ket{j}$. We begin with the following lemma.
	\begin{lem14}\label{lem14}
		If $\ket{R}\ne\ket{S_j}$ and $y_j\ne0$, then $x_j=x_{j+1\mmod{m}}$. Similarly, if $x_j\ne0$ and $\ket{R}\ne\ket{L_{jk}}$ for any $k$, then $y_i=y_{i^\prime}$ for all $i,i^\prime\ne j,j-1\mmod{m}$.
	\end{lem14}
	\proof Recall the definitions of $\ket{T_j},\ket{L_{j0}}$ given below \myeq{eqn3001} of the main text. When $\ket{R}\ne\ket{S_j}$, then from \myeq{eqn6001} we have that $x_jy_j=\inpd{jj}{\phi}=b_j/\sqrt{2}=\inpd{j+1,j}{\phi}=x_{j+1}y_j$. Therefore, if $y_j\ne0$ the first claim follows immediately. When $\ket{R}\ne\ket{L_{jk}}$, $x_jy_i=\inpd{ji}{\phi}=c_{j0}/\sqrt{n-2}=\inpd{ji^\prime}{\phi}=x_{j}y_{i^\prime}$ for all $i,i^\prime\ne j,j-1\mmod{m}$ and the second claim follows, completing the proof.\endproof
	
	\begin{lem12}\label{lem12}
		Excluding the case $m=3,n=4$, if for all $i,j$, $\inpd{ij}{\phi}\ne0$, then $x_i=x_0\ne0$ for all $i$ and $y_j=y_0\ne0$ for all $j$, which gives $\ket{\phi}\propto\ket{F}$ as the only product state orthogonal to all but one of the GenTiles$2$ states. Therefore, the omitted state cannot be $\ket{S_j}$ or $\ket{L_{jk}}$ but must be $\ket{F}$ itself. In case $m=3,n=4$, then the (unnormalized) state $\left(\ket{0}+\ket{1}+\ket{2}\right)\otimes\left(\sum_{i\ne j+1}\ket{i}-3\ket{j+1}\right)$ is orthogonal to all states in GenTiles$2$ other than  $\ket{L_{j1}}=\ket{j}\otimes\left(\ket{j+1\mmod{3}}-\ket{3}\right)/\sqrt{2}$.
	\end{lem12}
	\proof Since none of the $x_i,y_i$ vanish, we can invoke Lemma~\ref{lem14}. Even if $\ket{R}=\ket{S_J}$, then due to the circular relationship imposed by $\mmod{m}$, we have in succession, $x_{J+1}=x_{J+2},x_{J+2}=x_{J+3},\cdots,x_{m-2}=x_{m-1},x_{m-1}=x_0,\cdots, x_{J-2}=x_{J-1},x_{J-1}=x_J$, and all the $x_j$ are identical. Similarly, even if $\ket{R}=\ket{L_{JK}}$, there are always two non-adjacent columns that are not the $J^\textrm{th}$ (when $m>3$): $j-1,j+1\mmod{m}$, such that the second claim of Lemma~\ref{lem14} applies for each of these columns, from which we may conclude that all the $y_j$ are identical, which completes the proof for the general case.
	
	For the case, $m=3$, columns $j-1,j+1\mmod{m}$ are adjacent to each other. Therefore when $\ket{R}=\ket{L_{Jk}}$ for some $k$, row $J+1$ may be different from the others, those others still being equal to each other. That is, $y_i=y_j$ for all $i,j\ne J+1\mmod{m}$, while we still have $x_j=x_0$ for all $j$. With orthogonality to $\ket{F}$, this leaves us with a state $(\ket{0}+\ket{1}+\ket{2})\otimes(\sum_{j\ne J+1}\ket{j}-(n-1)\ket{J+1})$. However, this is not orthogonal to any of the states $\ket{L_{Jk}}$ for $k\ne0$. Therefore, this state is an acceptable solution only when $n=4$ and $\ket{R}=\ket{L_{J1}}$, which is the only state of GenTiles$2$ in that long tile, and this completes the proof.\endproof
	
	To find other acceptable product states, we may thus assume that $\phi_{ij}=\inpd{ij}{\phi}=0$ for at least one pair of indices, $i,j$. We begin with,
	\begin{lem13}\label{lem13}
		The number of zero entries in any given tile is either $0,1$ or equal to the length of the tile (short tiles have length $2$, long tiles $n-2$). Furthermore, there can be no more than one tile that has $1$ zero entry, that being the tile containing $\ket{R}$.
	\end{lem13}
	\proof Consider the $J^\textrm{th}$ short tile, whose entries are determined by $a_J\ket{S_J}+b_J\ket{T_J}$. If $\ket{R}\ne\ket{S_J}$, then orthogonality to $\ket{S_J}$ requires $a_J=0$. Then, we have no zero entries when $b_J\ne0$, and $2$ zero entries when $b_J=0$. On the other hand, when $\ket{R}=\ket{S_J}$ and $b_J=\pm a_J$, we have $1$ zero entry in that tile.
	
	For the $J^\textrm{th}$ long tile, we follow a similar argument. This tile's entries are determined by the coefficients $c_{Jk}$, where $c_{Jk}=0$ for all $k\ne0$ unless $\ket{R}\ne\ket{L_{JK}}$, in which case both $c_{J0}$ and $c_{JK}$ can be non-vanishing. If both vanish, we have $n-2$ zero entries in that tile; if one vanishes and the other does not, we have no zero entries; and finally, if both are non-zero and related as $c_{JK}=-\omega^{iK}c_{J0}$, we have a single zero entry in that tile. Clearly, the case of a single zero entry can only occur when $\ket{R}$ resides within that given tile, and this completes the proof.\endproof
	
	We will also use the following lemma.
	\begin{lem15}\label{lem15}
		If there is a long tile that has all its entries equal to zero but that column is not all zeros, then the only allowed non-vanishing product states are $\ket{S_j},\ket{T_j}$ (depending on which state is chosen for $\ket{R}$), unless $n=4$. When $n=4$, states $\ket{L_{j0}}+\ket{L_{j1}}-\ket{T_{j+1}}$ are also allowed when $m=3,4$; and 
		$\ket{L_{j0}}-\ket{L_{j1}}+\ket{T_{j+2}}$,
		% $\ket{L_{j+1,0}}-\ket{L_{j+1,1}}-\ket{T_{j-1}}$, $\ket{L_{j+2,0}}-\ket{L_{j+2,1}}-\ket{T_{j}}$, $\ket{L_{j+2,0}}+\ket{L_{j+2,1}}-\ket{T_{j-1}}$,
 		as well as $\ket{L_{j1}}$ and $\ket{L_{j0}}$, are also acceptable states when $m=4$. [All indices are to be understood as $\mmod{m}$.]
	\end{lem15}
	\proof The entries in the $J^\textrm{th}$ long tile are equal to $x_Jy_i$ for all $i\ne J-1,J$. Therefore, under the assumptions of the lemma, $y_i=0$ for all $i\ne J-1,J$, and there are no more than two non-zero rows. Given the length of each long tile is $n-2$, then when $n>5$, there are at least $n-2-2>1$ zero entries in each of the long tiles. According to Lemma~\ref{lem13}, this implies that every long tile is all zeros, and the only contribution to $\ket{\phi}$ are from the $\ket{S_j},\ket{T_j},j=J-1,J$. Then, it is easy to see that no linear combination of these is a product state, except $a_{J-1}\ket{S_{J-1}}+b_{J-1}\ket{T_{J-1}}$ and $a_J\ket{S_J}+b_J\ket{T_J}$. Given the constraint of orthogonality to all states of GenTiles$2$ but the one chosen as $\ket{R}$, we are left with the only allowed product states being $\ket{S_j},\ket{T_j},j=J-1,J$, depending on the choice of $\ket{R}$, which completes the proof for $n>5$.
	
	When $n=5$, there can be $n-2-2=1$ zero in one of the long blocks, if that block contains $\ket{R}$. Then, the only additional product state is a linear combination of $\ket{L_{jk}}$ and $\ket{L_{j0}}$ for that $j^\textrm{th}$ block (with both coefficients necessarily non-zero in order for the block to have that one zero entry), but this is not orthogonal to $\ket{F}$, so leads to no new acceptable states under these circumstances.
		
\begin{figure}
		\centering
		\begin{minipage}{2in}%
			\centering
			\includegraphics[scale=1.2]{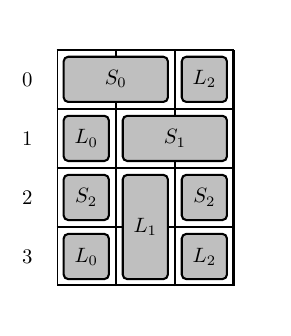}%\caption{\label{a}}
			\caption*{(a)}%
			\label{fig602a}%
		\end{minipage}%
		\qquad\qquad\qquad
		\begin{minipage}{2in}%
			\includegraphics[scale=1.2]{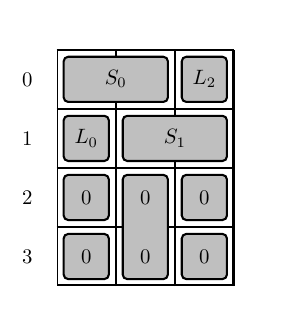}%\caption{\label{b}}
			\caption*{(b)}%
			\label{fig602b}%
		\end{minipage}%
		\caption{\label{fig602} GenTiles$2$ on a $3\times4$ system. Each tile is labeled in part (a). In part (b), the case where the $L_1$ tile is all zeros but $x_1\ne0$ is depicted, with zeros indicated in the bottom two rows, representing $y_2=0=y_3$; see Lemma~\ref{lem15}. $\ket{\phi}=\ket{L_{21}}+\ket{L_{20}}-\ket{T_{0}}$ lies entirely in row $0$ so is a product state, while $\ket{\phi}=\ket{L_{01}}+\ket{L_{00}}-\ket{T_{1}}$ lies entirely in row $1$ so is also product.}
	\end{figure}

\begin{figure}
		\centering
		\begin{minipage}{2.5in}%
			\centering
			\includegraphics[scale=1.2]{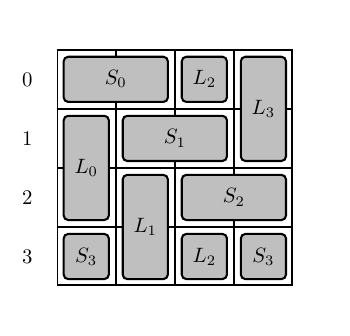}
			\caption*{(a)}%
			\label{fig603a}%
		\end{minipage}%
		\qquad\qquad\qquad
		\begin{minipage}{2.6in}%
			\includegraphics[scale=1.2]{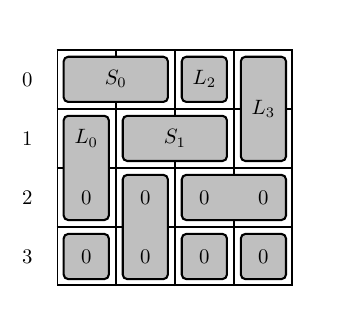}%
			\caption*{(b)}%
			\label{fig603b}%
		\end{minipage}%
		\caption{\label{fig603} GenTiles$2$ on a $4\times4$ system. Each tile is labeled in part (a). In part (b), the case where the $L_1$ tile is all zeros but $x_1\ne0$ is depicted, with zeros indicated in the bottom two rows, representing $y_2=0=y_3$; see Lemma~\ref{lem15}. $\ket{\phi}=\ket{L_{20}}-\ket{L_{21}}-\ket{T_{0}}$ lies entirely in row $0$, as does $\ket{\phi}=\ket{L_{30}}+\ket{L_{31}}-\ket{T_{0}}$, so both are product states, while $\ket{\phi}=\ket{L_{00}}+\ket{L_{01}}-\ket{T_{1}}$ and $\ket{\phi}=\ket{L_{30}}-\ket{L_{31}}-\ket{T_{1}}$ both lie entirely in row $1$ so are also product. In addition, $\ket{L_{30}}$ ($\ket{L_{31}}$) is also an acceptable state when $\ket{R}$ is chosen as $\ket{F}$ ($\ket{L_{31}}$).} 
	\end{figure}

	When $n=4$, the long blocks have length $2$ and can either fit entirely within the two non-zero rows, or can have one zero entry outside those rows and one non-zero entry inside them. When $m=3$ (see Figure~\ref{fig602}), each long tile intersects row $3$, so we have $y_3=0$ in this case and every long tile has a zero in it. One of these tiles can have only one zero ($L_0$ or $L_2$ in Figure~\ref{fig602}), but the other must be all zeros. Then, in addition to the acceptable product states already identified, we also find $\ket{L_{J+1,1}}+\ket{L_{J+1,0}}-\ket{T_{J+2}}$ and $\ket{L_{J-1,1}}+\ket{L_{J-1,0}}-\ket{T_{J}}$ are product states orthogonal to all the states in GenTiles$2$ except $\ket{L_{J\pm1,1}}$, respectively (and all subscripts are mod $m$, so each of these product states is of the form indicated in the lemma). When $m=4$ (see Figure~\ref{fig603}), a similar kind of situation arises and here we obtain $\ket{L_{J-1,0}}+\ket{L_{J-1,1}}-\ket{T_{J}}$, $\ket{L_{J+1,0}}-\ket{L_{J+1,1}}-\ket{T_{J-1}}$, $\ket{L_{J+2,0}}-\ket{L_{J+2,1}}-\ket{T_{J}}$, $\ket{L_{J+2,0}}+\ket{L_{J+2,1}}-\ket{T_{J-1}}$ when $\ket{R}$ is the given $\ket{L_{j1}}$ (again, each of these is of the form indicated in the lemma, with mod $m$). In addition, here, a long tile can be without any zero entries, so we also have $\ket{L_{j1}}$ and $\ket{L_{j0}}$ as acceptable states, depending on the choice of $\ket{R}$. This completes the proof.\endproof
	
	We can now complete our search for allowable product states by considering the follwowing case: if for every $j$ such that the $j^\textrm{th}$ long tile is all zeros, then $x_j=0$, see Lemma~\ref{lem17} below. First, we have
	\begin{lem16}\label{lem16}
		If there exists pair $i,j$ such that $\inpd{ij}{\phi}=0$, then there also exists $s$ such that $y_s=0$.
	\end{lem16}	
	\proof The proof is by contradiction, so assume $x_iy_j=\inpd{ij}{\phi}=0$, but for all $s$, $y_s\ne0$. Then $x_i=0$ and the $i^\textrm{th}$ and $(i-1)^\textrm{th}$ short tiles each have a zero in them. It must be that $\ket{R}\ne\ket{S_i}$ or $\ket{R}\ne\ket{S_{i-1}}$, so starting with the one that is not $\ket{R}$, apply the argument used in the proof of Lemma~\ref{lem12}. For example, suppose $\ket{R}\ne\ket{S_i}$. Then, the other entry in the $i^\textrm{th}$ short tile, which is $x_{i+1}y_i$ is also zero. Since by assumption, $y_i\ne0$, we have that $x_{i+1}=0$, implying that the $(i+1)^\textrm{th}$ short tile is all zeros, so that also, $x_{i+2}y_{i+1}=0$. Therefore, $x_{i+2}=0$, and by continuing the argument around the circle (mod $m$), we find $x_l=0$ for all $l$ and we have no non-zero state under these conditions. If one of the other $\ket{S_l}$ is the chosen $\ket{R}$, then one can go around in both directions starting at $x_{i-1},x_i$ to finish at $x_l$ coming from both sides, arriving at the same conclusion. This completes the proof.\endproof
	
\begin{figure}
		\centering
		\includegraphics[scale=1.2]{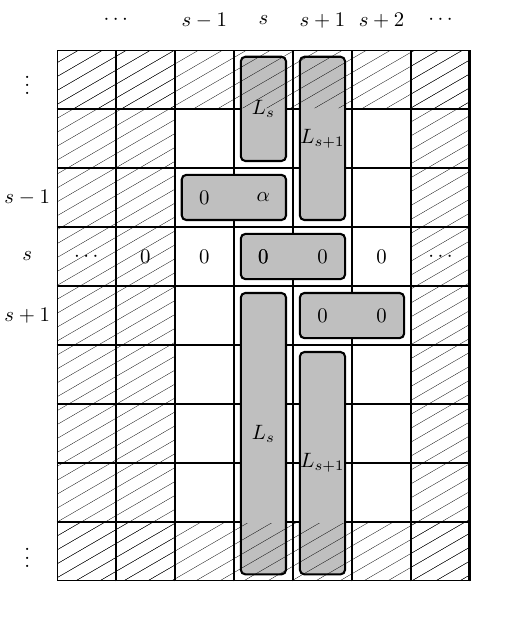}%\caption{\label{a}}
		\caption{\label{fig604} Depicted here is the case where $y_s=0$ and $\ket{R}=\ket{F}$ or $\ket{S_j}$. The coefficient $\alpha$ shown in row $s-1$, column $s$, must vanish unless $\ket{R}=\ket{S_{s-1}}$.} 
	\end{figure}

	The last step is
		\begin{lem17}\label{lem17}
		If every long tile that is all zeros sits in a column that is also all zeros, then the only allowable product states under these circumstances are $\ket{L_{j0}}$ (if $\ket{R}=\ket{F}$) and $\sqrt{2}\ket{L_{j+1,0}}-\sqrt{n-2}\left(\ket{T_{j}}-\ket{S_{j}}\right)$ and $\sqrt{2}\ket{L_{j0}}-\sqrt{n-2}\left(\ket{T_{j}}+\ket{S_{j}}\right)$ (the latter two apply when $\ket{R}=\ket{S_j}$).
	\end{lem17}	
	\proof Since from Lemma~\ref{lem16}, at least one row is all zeros, there are no more than two of the long tiles that do not have a zero in them. Let us first consider the case where $\ket{R}=\ket{F}$ or $\ket{S_j}$ for some $j$. In this case, for which the long tiles have only $\ket{L_{j0}}$ in them, one zero entry implies that the entire long tile is zero. Under the conditions of the present Lemma, this means there are no more than two columns that have non-zero entries in them. This situation, with $y_s=0$ leaving the $s^\textrm{th}$ and $(s+1)^\textrm{th}$ columns the only ones that aren't all zero, is depicted in Figure~\ref{fig604}. Note that $\alpha\ne0$ only if $\ket{R}=\ket{S_{s-1}}$.
	
	When $\ket{R}=\ket{F}$, $\alpha=0$ in Figure~\ref{fig604} and the only contributions to $\ket{\phi}$ are from $\ket{L_{s0}}$ and $\ket{L_{s+1,0}}$. It is clear that the only linear combination of these two states that yields a product state are the individual states $\ket{L_{s0}}$ and $\ket{L_{s+1,0}}$, each by itself.
	
	When $\ket{R}=\ket{S_j}$ with $j\ne s-1$, we still require $\alpha=0$, and we are left with the same situation as for $\ket{R}=\ket{F}$. However, we here require orthogonality to $\ket{F}$, so we find no allowable states in this case.
	
	When $\ket{R}=\ket{S_{s-1}}$, we have $\ket{\phi}=c_{s0}\ket{L_{s0}}+c_{s+1,0}\ket{L_{s+1,0}}+b_{s-1}\ket{T_{s-1}}+a_{s-1}\ket{S_{s-1}}$. Here, the zero appearing in the short tile at position $(s+1,s+1)$ requires that either $c_{s0}=0$ (from $y_{s+1}=0$) or $c_{s+1,0}=0$ (from $x_{s+1}=0$). If $c_{s+1,0}\ne0$ here, then also $\alpha=0$ to obtain a product state, and we have $\ket{L_{s+1,0}}$, which is not allowed due to the requirement of orthogonality to $\ket{F}$. We do find an allowable state when $c_{s+1,0}=0$, that being $\sqrt{2}\ket{L_{s0}}-\sqrt{n-2}\left(\ket{T_{s-1}}-\ket{S_{s-1}}\right)$. Similarly, when $\ket{R}=\ket{S_{s+1,0}}$, we obtain the allowable state, $\sqrt{2}\ket{L_{s+1,0}}-\sqrt{n-2}\left(\ket{T_{s+1}}+\ket{S_{s+1}}\right)$.
	
	We are left to consider the case where $\ket{R}=\ket{L_{jk}}$ for some $j,k$. In the case $j\ne s-1$, we must have $\alpha=0$ again in Figure~\ref{fig604}, and the only contributions are from the two long tiles depicted in the figure, along with (if also $j\ne s,s+1$) additional contributions from the $j^\textrm{th}$ long tile. Then, $\ket{\phi}=c_{jk}\ket{L_{jk}}+c_{j0}\ket{L_{j0}}+c_{s0}\ket{L_{s0}}+c_{s+1,0}\ket{L_{s+1,0}}$, and it is easy to see that no product state orthogonal to $\ket{F}$ is possible in this case (whether or not $j=s,s+1$, in which case the second appearance of $\ket{L_{s0}}$ or $\ket{L_{s+1,0}}$ should obviously be omitted).
	
	Finally, if $j=s-1$, then in Figure~\ref{fig604}, the zero in that short tile next to $\alpha$ should be replaced by another instance of $\alpha$, which need not vanish. Now we have  $\ket{\phi}=c_{s-1,0}\left(\ket{L_{s-1,0}}-\ket{L_{s-1,1}}\right)+c_{s0}\ket{L_{s0}}+c_{s+1,0}\ket{L_{s+1,0}}+b_{s-1}\ket{T_{s-1}}$ (this is the combination that makes the entry in the $s^\textrm{th}$ row within the $(s-1)^\textrm{th}$ long tile vanish, as it must since $y_s=0$). Noting in this altered version of Figure~\ref{fig604} that the $(s-1,s-2)$ entry, which is equal to $x_{s-1}y_{s-2}$, must vanish (it is part of the $(s-2)^\textrm{th}$ short tile), then if $x_{s-1}\ne0$, $y_{s-2}=0$ which requires $c_{s0}=0=c_{s+1,0}$. Then the only product states are either $\ket{T_{s-1}}$ or $\ket{L_{s-1,0}}-\ket{L_{s-1,k}}$, neither of which is orthogonal to $\ket{F}$, so are not allowed. If, on the other hand, $x_{s-1}=0$, then we have a situation already considered in preceding paragraphs. This completes the proof.\endproof
	
	Collecting all of the results in this appendix, we obtain Lemma~\ref{lem11}.
	
\subsection{Complete separable measurements for discriminating GenTiles$2$}\label{AppB2}
Any acceptable measurement consists of operators proportional to projectors onto the states listed in Lemma~\ref{lem11}. A complete measurement requires that a positive linear combination of these operators is equal to the identity, see \myeq{eqn100} in the main text. When $n>4$, let us list the various allowable states as
\begin{align}\label{eqn7001}
	\ket{\Psi_0}&=\ket{F}\notag\\
	\ket{\Psi_{j+1}}&=\ket{S_j}\notag\\
	\ket{\Psi_{m+j+1}}&=\ket{T_j}\notag\\
	\ket{\Psi_{(k+2)m+j+1}}&=\ket{L_{jk}}\notag\\
	\ket{\Psi_{nm+j+1}}&=\sqrt{n-2}\ket{L_{j0}}-\frac{n-2}{\sqrt{2}}\left(\ket{T_j}+\ket{S_j}\right)\notag\\
					&=\ket{j}\otimes\left(\sum_{i=0}^{m-3}\ket{i+j+1\mmod{m}}+\sum_{i=m}^{n-1}\ket{i}-(n-2)\ket{j}\right)\notag\\
	\ket{\Psi_{(n+1)m+j+1}}&=\sqrt{n-2}\ket{L_{j0}}-\frac{n-2}{\sqrt{2}}\left(\ket{T_{j-1}}-\ket{S_{j-1}}\right)\notag\\
	&=\ket{j}\otimes\left(\sum_{i=0}^{m-3}\ket{i+j+1\mmod{m}}+\sum_{i=m}^{n-1}\ket{i}-(n-2)\ket{j-1}\right).
\end{align}
The projectors onto these states are all diagonal in the standard basis on one or the other party, except for the one proportional to $[F]$. Taking matrix elements of \myeq{eqn100},
\begin{align}\label{eqn7002}
	\bra{00}\left(\sum_jc_j[\Psi_j]\right)\ket{11}=\bra{00}I_\HC\ket{11}
\end{align}
reduces to $c_0=0$, implying that $[F]$ must be excluded from the measurement. Given this, next take
\begin{align}\label{eqn7003}
	\bra{jj}\left(\sum_jc_j[\Psi_j]\right)\ket{j,n-1}=0,
\end{align}
and
\begin{align}\label{eqn7004}
	\bra{j,j-1}\left(\sum_jc_j[\Psi_j]\right)\ket{j,n-1}=0,
\end{align}
which gives $c_{nm+j+1}=0$ and $c_{(n+1)m+j+1}=0$, respectively, for $j=0,\cdots,m-1$. This leaves us with a unique complete separable measurement for discriminating GenTiles$2$ when $n>4$, that consisting of projectors onto the states, $\ket{S_j},\ket{T_j}$, and $\ket{L_{jk}}$, for $j=0,\cdots,m-1$ and $k=0,\cdots,n-3$. This completes the proof of Lemma~\ref{lem18} for $n>4$.

In the case of $m=n=4$, the allowed states are
\begin{align}\label{eqn7005}
	\ket{\Psi_0}&=\ket{F}=\left(\ket{0}+\ket{1}+\ket{2}+\ket{3}\right)\otimes\left(\ket{0}+\ket{1}+\ket{2}+\ket{3}\right)\notag\\
	\ket{\Psi_{j+1}}&=\ket{S_j}=\left(\ket{j}-\ket{j+1\mmod{3}}\right)\otimes\ket{j}\notag\\
	\ket{\Psi_{j+5}}&=\ket{T_j}=\left(\ket{j}+\ket{j+1\mmod{3}}\right)\otimes\ket{j}\notag\\
	\ket{\Psi_{3k+j+9}}&=\ket{L_{jk}}=\ket{j}\otimes\left(\ket{j+1\mmod{4}}+(-1)^k\ket{j+2\mmod{4}}\right)\notag\\	\ket{\Psi_{j+17}}&=\ket{L_{j0}}-\left(\ket{T_j}+\ket{S_j}\right)\notag\\
					&=\ket{j}\otimes\left(\ket{j+1\mmod{4}}+\ket{j+2\mmod{4}}-2\ket{j}\right)\notag\\
	\ket{\Psi_{j+21}}&=\ket{L_{j0}}-\left(\ket{T_{j-1}}-\ket{S_{j-1}}\right)\notag\\
					&=\ket{j}\otimes\left(\ket{j+1\mmod{4}}+\ket{j+2\mmod{4}}-2\ket{j-1\mmod{4}}\right)\notag\\
	\ket{\Psi_{j+25}}&=\ket{T_j}-\left(\ket{L_{j-1,0}}+\ket{L_{j-1,1}}\right)\notag\\
					&=\left(\ket{j}+\ket{j+1\mmod{4}}-2\ket{j-1\mmod{4}}\right)\otimes\ket{j}\notag\\
	\ket{\Psi_{j+29}}&=\ket{T_j}-\left(\ket{L_{j-2,0}}-\ket{L_{j-2,1}}\right)\notag\\
					&=\left(\ket{j}+\ket{j+1\mmod{4}}-2\ket{j-2\mmod{4}}\right)\otimes\ket{j},
\end{align}
with $j=0,1,2,3$ and $k=0,1$. First note that $\ket{F}$ is the only state that is non-diagonal in the standard basis on both parties. Therefore, taking the $\bra{i_1i_2}\cdots\ket{i_1^\prime i_2^\prime}$ matrix element of \myeq{eqn100} gives that $c_0=0$ again, and $\ket{F}$ cannot be a part of the measurement. Then, taking the $\bra{jj}\cdots\ket{j,j+1}$ matrix element of \myeq{eqn100} shows that $c_{j+17}=0$ for all $j$. Taking the $\bra{jj}\cdots\ket{j,j+2}$ matrix element of \myeq{eqn100} shows that $c_{j+21}=0$ for all $j$. Similarly, the matrix elements $\bra{jj}\cdots\ket{j-1,j}$ and $\bra{jj}\cdots\ket{j-2,j}$ show that $c_{j+25}=0$ and $c_{j+29}=0$, respectively. This again leaves the only operators allowed in the complete measurement as those stated in Lemma~\ref{lem18}. This completes the proof for $m=n=4$.

The final case of $m=3,n=4$ is slightly more involved algebraically. The allowed states are
\begin{align}\label{eqn7006}
	\ket{\Psi_0}&=\ket{F}=\left(\ket{0}+\ket{1}+\ket{2}\right)\otimes\left(\ket{0}+\ket{1}+\ket{2}+\ket{3}\right)\notag\\
	\ket{\Psi_{j+1}}&=\ket{S_j}=\left(\ket{j}-\ket{j+1\mmod{3}}\right)\otimes\ket{j}\notag\\
	\ket{\Psi_{j+4}}&=\ket{T_j}=\left(\ket{j}+\ket{j+1\mmod{3}}\right)\otimes\ket{j}\notag\\
	\ket{\Psi_{3k+j+7}}&=\ket{L_{jk}}=\ket{j}\otimes\left(\ket{j+1\mmod{3}}+(-1)^k\ket{3}\right)\notag\\
	\ket{\Psi_{j+13}}&=\ket{L_{j0}}-\left(\ket{T_j}+\ket{S_j}\right)\notag\\
	&=\ket{j}\otimes\left(\ket{j+1\mmod{3}}+\ket{3}-2\ket{j}\right)\notag\\
	\ket{\Psi_{j+16}}&=\ket{L_{j0}}-\left(\ket{T_{j-1}}-\ket{S_{j-1}}\right)\notag\\
	&=\ket{j}\otimes\left(\ket{j+1\mmod{3}}+\ket{3}-2\ket{j-1\mmod{3}}\right)\notag\\
	\ket{\Psi_{j+19}}&=\ket{T_j}-\left(\ket{L_{j-1,0}}+\ket{L_{j-1,1}}\right)\notag\\
	&=\left(\ket{j}+\ket{j+1\mmod{3}}-2\ket{j-1\mmod{3}}\right)\otimes\ket{j}\notag\\
	\ket{\Psi_{j+22}}&=\left(\ket{0}+\ket{1}+\ket{2}\right)\otimes\left(3\ket{j}-\sum_{i\ne j}^3\ket{i}\right),
\end{align}
with $j=0,1,2$ and $k=0,1$. From the $\bra{00}\cdots\ket{11}$, $\bra{00}\cdots\ket{12}$, $\bra{00}\cdots\ket{13}$, and $\bra{01}\cdots\ket{12}$ matrix elements of \myeq{eqn100}, only $\ket{F}$ and $\ket{\Psi_{j+22}}$ contribute and we obtain in turn,
\begin{align}\label{eqn7007}
	0&=c_{0}-3c_{22}-3c_{23}+c_{24},\notag\\
	0&=c_{0}-3c_{22}+c_{23}-3c_{24},\notag\\
	0&=c_{0}-3c_{22}+c_{23}+c_{24},\notag\\
	0&=c_{0}+c_{22}-3c_{23}-3c_{24}.
\end{align}
\noindent The determinant of the matrix of coefficients for this set of equations is equal to $64$, which implies that the only solution is for each of the $c_j$ appearing in that equation must vanish. Therefore, $\ket{F}$ and states $\ket{\Psi_{j+22}}$ must be omitted from the measurement. Excluding these states, consider the $\bra{0j}\cdots\ket{1j}$ matrix elements of \myeq{eqn100} for $j=1,2$. This yields $c_{20}=0$ from $j=1$, $c_{21}=0$ from $j=2$. Then, taking the $\bra{00}\cdots\ket{20}$ matrix element gives $c_{19}=0$, as well, and states $\ket{\Psi_{j+19}}$ are excluded. Finally, from matrix elements $\bra{j0}\cdots\ket{j1}$ for $j=0,2$, $\bra{j0}\cdots\ket{j2}$ for $j=1,2$, and $\bra{j1}\cdots\ket{j2}$ for $j=0,1$, we obtain in succession that $c_{13},c_{18},c_{17},c_{15},c_{16},c_{14}$ each must vanish. We are again left with only the states $\ket{S_j},\ket{T_j},\ket{L_{jk}}$ as claimed in Lemma~\ref{lem18}, and this completes the proof.\endproof

\section{Non-existence of paths of product operators from $I_\HC$ to the outcomes of $\MC_{GT1}$ for GenTiles$1$}\label{AppA3}
We now prove that the measurement $\MC_{GT1}$ of Lemma~\ref{lem6} is not in $\locc$. We will characterize all product operators that lie in the zonotope generated by the outcomes of $\MC_{GT1}$, those being the projectors $\{[H_{km}]\}$, $\{[V_{km}]\}$, denoting this zonotope as $\ZC_{\MC_{GT1}}$. We will use the following lemma in what follows.
\begin{lem10}\label{lem10}
	A positive linear combination of projectors onto the states in one of the $H$-tiles (or in one of the $V$-tiles) leads to a matrix whose entries are all zeros except for an $n/2\times n/2$ block along the diagonal that is of the Toeplitz form in the standard basis. This Toeplitz block is diagonal if and only if the positive linear combination is proportional to a simple sum of the projectors onto the given $H$-tile ($V$-tile) states, and is then itself proportional to a projector of rank $n/2$.
\end{lem10}
\proof Consider a positive linear combination of projectors onto the states in one of the $H$-tiles, say $H_k$. This has $B$-part equal to projector $[k]$, and $A$-part
\begin{align}\label{eqn1113}
	\hat\AC^{(k)}&=\sum_{m=0}^{\frac{n}{2}-1}\hat c_{km}\sum_{i,j=0}^{n/2-1}\omega^{m(j-i)}\ket{j+k\mmod{n}}\bra{i+k\mmod{n}}\notag\\
	&=\sum_{i,j=0}^{n/2-1}\tilde c_{k,j-i}\ket{j+k\mmod{n}}\bra{i+k\mmod{n}},
\end{align}
where $\tilde c_{k,l}=\sum_{m=0}^{n/2-1}\hat c_{km}\omega^{ml}=\tilde c_{k,-l}^\ast$. Since the coefficients depend only on $j-i$, this has an $n/2\times n/2$ block along the diagonal that has the Toeplitz form,
\begin{align}\label{eqn1015}
	\begin{bmatrix}
		\tilde c_{k,0}&\tilde c_{k,1}&\tilde c_{k,2}&\tilde c_{k,3}&\cdots&\\
		\tilde c_{k,-1}&\tilde c_{k,0}&\tilde c_{k,1}&\tilde c_{k,2}&\ddots&\\
		\tilde c_{k,-2}&\tilde c_{k,-1}&\tilde c_{k,0}&\tilde c_{k,1}&\ddots&\\
		\tilde c_{k,-3}&\tilde c_{k,-2}&\tilde c_{k,-1}&\tilde c_{k,0}&\ddots&\\
		\vdots&\ddots&\ddots&\ddots&\ddots\\
	\end{bmatrix}
\end{align}
Furthermore, $\tilde c_{k,0}=0$ if and only if $\hat c_{km}=0$ for all $m$, and then $\hat\AC^{(k)}=0$. In addition, this block is diagonal if and only if $\tilde c_{k,r}=0$ for all $r\ne0$, implying that  the vector of coefficients $\hat c_{km}$ is orthogonal to each of the vectors $\vec\omega_r$, with coefficients $(\vec\omega_r)=\omega^{mr}$, in which case $\hat c_{km}=\hat c_k$, independent of $m$. Then, $\hat\AC^{(k)}$ is diagonal and proportional to a rank-$n/2$ projector, and this completes the proof.\endproof

\noindent Note that as $k$ is incremented, these Toeplitz blocks shift one entry to the right and downward within the full $n\times n$ matrix $\AC$.

In order to prove Theorem~\ref{thm9}, we need to show that no continuous path of product operators from $I_\HC$ to at least one of the outcomes of a complete measurement distinguishing the states of GenTiles$1$ exists. We will see that no such path exists to any of the outcomes in the measurement found above in Lemma~\ref{lem6}. Referring to Lemma~\ref{lem6}, which tells us which projectors are to be included in our complete separable measurement, we will next find all product operators that are positive linear combination of those projectors, and so must have the form
\begin{align}\label{eqn1110}
	\AC\otimes\BC=\sum_{k=0}^{n-1}\sum_{m=0}^{\frac{n}{2}-1}\left(c_{km}[H_{km}]+c_{km}^\prime[V_{km}]\right),
\end{align}
with non-negative coefficients. We next prove the following lemma, in which we denote the standard basis on either party by `SB'.
\begin{lem2}\label{lem2}
	If $\AC$ is not diagonal in SB, then $\BC$ is diagonal in SB and $\AC\otimes\BC$ has rank no greater than $n/2$. The same conclusion holds when the roles of the two parties is exchanged.
\end{lem2}
\proof Since $[V_{km}]$ is diagonal in SB on the $A$ side, then for $\kappa^\prime\ne\kappa$,
\begin{align}\label{eqn1011}
	\AC_{\kappa\kappa^\prime}\BC&=\sum_{k=0}^{n-1}\sum_{m=0}^{\frac{n}{2}-1}c_{km}~_A\inpd{\kappa}{H_{km}}\inpd{H_{km}}{\kappa^\prime}_A.
\end{align}
Consider the case $\kappa^\prime>\kappa$; since $\AC$ is Hermitian, this effectively covers all cases. Since the $H_k$ tile stretches from $k$ to $k+n/2-1\mmod{n}$ in the $A$-space, this constrains which $H$-tiles contribute in \myeq{eqn1011} for given $\kappa,\kappa^\prime$. Notice in particular, that if $\kappa^\prime-\kappa=n/2$, then $\AC_{\kappa\kappa^\prime}=0$ since every term in the sum on the right-hand side of \myeq{eqn1011} vanishes. More generally, the only values of $k$ that contribute in this equation are those such that both $\kappa$ and $\kappa^\prime$ lie in the range from $k$ to $k+n/2-1\mmod{n}$. Introducing $\mu=\kappa-n/2+1\mmod{n}$ and $\mu^\prime=\kappa^\prime-n/2+1\mmod{n}$, and recalling the definition of $\tilde c_{kl}$ given below \myeq{eqn1113}, we have
\begin{align}\label{eqn1112}
	\AC_{\kappa\kappa^\prime}\BC=
	\begin{cases}
		~\sum\limits_{k=\mu^\prime}^{\kappa}\tilde c_{k,\kappa-\kappa^\prime}[k]~~~\kappa^\prime-\kappa=1,2,\cdots,n/2-1\vspace{2mm}\\
		~\sum\limits_{k=\mu}^{\kappa^\prime}\tilde c_{k,\kappa-\kappa^\prime}[k]~~~\kappa^\prime-\kappa=n/2+1,\cdots,n-1  
	\end{cases}
\end{align}
and then rank$(\BC)\le n/2$. This tells us that if $\AC$ is not diagonal in SB on the $A$-side, then $\BC$ is diagonal in SB on the $B$-side. (Since the set of states are symmetric under exchange of the parties, we also have that when $\BC$ is not diagonal, then $\AC$ is diagonal, again in SB on each side.)

Since $\BC$ is diagonal in SB, then according to Lemma~\ref{lem10}, there can be no contributions from the $V$-tiles at all, unless they are such that the corresponding Toeplitz blocks (for party $B$, here) are diagonal. This means that the only contributions from the $V$-tiles must be of the form of rank-$n/2$ projectors onto whichever $V$-tile is contributing. However, since we have seen that $\BC$ has rank no greater than $n/2$, and since each of these projectors onto a $V$-tile has support strictly different than that for any other $V$-tile, this is a contradiction if there is more than one $V$-tile that contributes. Therefore, we can assume that no more than one $V$-tile contributes, and that one contributing $V$-tile must be the one that has support matching that of $\BC$ shown in \myeq{eqn1112}. But this means that when $\AC$ is not diagonal in SB, $\AC\otimes \BC$ of \myeq{eqn1110} reduces to
\begin{align}\label{eqn1021}
	\AC\otimes\BC&=c_{\kappa0}^\prime[\kappa]\otimes\sum_{k=\kappa+1}^{\kappa+n/2}[k]+\sum_{k=0}^{n-1}\sum_{i,j=0}^{n/2-1}\tilde c_{k,j-i}\ket{j+k\mmod{n}}\bra{i+k\mmod{n}}\otimes[k],
\end{align}
where the first term on the right is from the single contributing $V_\kappa$-tile. Now, in order for this to be a product operator, the $A$-parts of the various terms must all be proportional to each other for different $k$. However, the second term with the $i,j$ sum, which is of the Toeplitz form discussed in Lemma~\ref{lem10}, differs from each $k$ to the next by a shift of the Toeplitz block down and to the right, as noted just after the proof of that lemma. The first term $c_{\kappa0}^\prime[\kappa]$ cannot correct for that shift to make these expressions proportional to each other, so this is not a product operator unless all of the Toeplitz blocks vanish except for one. Then, the remaining Toeplitz block is tensored with a single $[k]$, whereas the $c_{\kappa0}^\prime[\kappa]$ term is tensored with a sum over $n/2$ different projectors $[k]$. Therefore, we must have $c_{\kappa0}^\prime=0$, since we assume here that $\AC$ is not diagonal in SB, and we are left with $\BC$ having rank equal to unity, in which case $\AC\otimes\BC$ has rank no more than $n/2$. These cases are linear combinations of projectors onto the states $\ket{H_{km}}$ for one fixed $k$. By the symmetry under exchange of the parties, then for the case where $\BC$ is not diagonal in SB, we also obtain linear combinations of projectors onto the states $\ket{V_{km}}$ for one fixed $k$, which again are of rank no greater than $n/2$. This completes the proof.\endproof

We are left to consider the case that $\AC$ and $\BC$ are each diagonal in their respective SB. We next prove our final lemma.
\begin{lem3}\label{lem3}
	The identity operator $I_\HC$, along with those proportional to it, are the only product operators $\AC\otimes\BC$ in $\ZC_\MC$ that has rank greater then $n/2$, and therefore, $(0,I_\HC]$ is an isolated line segment in the intersection of $\ZC_\MC$ with the set of product operators.
\end{lem3}
\proof We have already shown that product operators in $\ZC_\MC$ that are not diagonal in SB have rank no greater than $n/2$. Therefore, we need to consider those product operators that are diagonal. For $\kappa\ne\kappa^\prime$, then from \myeq{eqn1112},
\begin{align}\label{eqn1016}
	0=\AC_{\kappa\kappa^\prime}\BC_{kk}&=\tilde c_{k,\kappa-\kappa^\prime}.
\end{align}
For each $k$ and pairs $\kappa,\kappa^\prime$ within the range of the $H_k$-tile, $\lvert\kappa-\kappa^\prime\rvert$ ranges over all values from $1$ to $n/2-1$, modulo $n/2$ (using modulo $n/2$ here because $\omega^{n/2}=1$). Therefore, $\tilde c_{k,\kappa-\kappa^\prime}=0$ here for all $k$ and all $\kappa\ne\kappa^\prime$, which implies that $c_{km}=c_{k0}$ for all $m$ and all $k$. By the same argument but looking at $\AC_{kk}\BC_{\kappa\kappa^\prime}=0$ for $\kappa\ne\kappa^\prime$, we see also that $c_{km}^\prime=c_{k0}^\prime$ for those terms in \myeq{eqn1110} involving the $V$-tiles. This means that $\AC\otimes\BC$ is a (positive) linear combination of rank-$n/2$ projectors, each of which projects onto the (entire) support of one of the $H$-tiles, or onto that of one of the $V$-tiles. These projectors are each of the form
\begin{align}\label{eqn1119}
	P_H(k)&=\sum_{j=0}^{n/2-1}[j+k\mmod{n}]\otimes[k]\notag\\
	P_V(k)&=[k]\otimes\sum_{j=0}^{n/2-1}[j+k+1\mmod{n}].
\end{align}
We wish to find all linear combinations of these projectors that are product operators. Introducing the isomorphism, $[i]\mapsto\ket{i}$, these map as $P_H(k)\mapsto\ket{H_{k0}}$ and $P_V(k)\mapsto\ket{V_{k0}}$, and our present problem maps to the problem of finding all product states that are linear combinations of the $\ket{H_{k0}}$ and $\ket{V_{k0}}$. However, this is the same problem as finding all product states that are orthogonal to all the states of GenTiles$1$ other than $\ket{F}$, which we have already solved above. The answer found there---see the paragraph just above \myeq{eqn1119}---translates back to the problem here as $\{P_H(k)\}\mapsfrom\{\ket{H_{k0}}\}$, $\{P_V(k)\}\mapsfrom\{\ket{V_{k0}}\}$, and $I_\HC\mapsfrom\ket{F}$. The first two sets consist of product operators each of which has rank equal to $n/2$, while $I_\HC$ has rank of $n^2$, and the proof is complete.\endproof

A direct consequence of Lemma~\ref{lem3} is that there exists no continuous path of product operators lying within $\ZC_\MC$ and stretching from $I_\HC$ to anywhere. This completes the proof of Theorem~\ref{thm9} that GenTiles$1$ cannot be perfectly discriminated within $\locc$.

\section{Proof of Theorem~\ref{thm11}}\label{AppC}
Let $\omega=e^{2\pi i/N}$ and
\begin{align}\label{eqn24}
	\RC := \sum_{j=1}^N c_j \Psi_j = \frac{1}{D}\sum_{j=1}^N c_j\sum_{ m_1, m_1^\prime=0}^{d_1-1}\sum_{ m_2, m_2^\prime=0}^{d_2-1}\cdots\sum_{ m_P, m_P^\prime=0}^{d_P-1}\omega^{\left[j\sum_{\alpha=1}^Pp_\alpha\left( m_\alpha- m_\alpha^\prime\right)\right]}\ket{ m_1, m_2,\cdots, m_P}\bra{ m_1^\prime, m_2^\prime,\cdots, m_P^\prime},
\end{align}
and we wish to determine the conditions under which $\RC$ is a product operator of the form $\AC\otimes\bar\AC$, where $\bar\AC=\BC\otimes\CC\otimes\cdots$. From here on, we replace $\bar\AC$ by $\BC$. We will show that $\RC= \AC\otimes\BC$ if and only if either (1) $\RC=c_i\Psi_i$ for some fixed $i$ or (2) $\RC=cI_A\otimes I_B$, which occurs when $c_j=Dc/N$, independent of $j$. We begin by examining the structure of $\RC$, as given in \myeq{eqn24}.

First, let $D_2=D/d_1$ and notice that we can write
\begin{align}\label{eqn25}
	\RC = \frac{1}{D}\sum_{j=1}^N c_j\sum_{m_1,m_2=0}^{d_1-1}\sum_{n_1,n_2=0}^{D_2-1}\omega^{j\left[m_1-m_2+d_1\left(n_1-n_2\right)\right]}\ket{m_1,n_1}\bra{m_2,n_2},
\end{align}
where $n_1= m_2 + d_2\left( m_3 + d_3\left( m_4 + \cdots + d_{P-2}\left( m_{P-1}+ d_{P-1} m_P\right)\cdots\right)\right)$ and similarly for $n_2$. Then, if $\RC$ is a product operator, we have that
\begin{align}\label{eqn26}
	\bra{m_1,n_1}\RC\ket{m_2,n_2} = \bra{m_1}\AC\ket{m_2}\bra{n_1}\BC\ket{n_2} = \frac{1}{D}\sum_{j=1}^N c_j\omega^{j\left[m_1-m_2+d_1\left(n_1-n_2\right)\right]}.
\end{align}
Defining $s_r=\sum_jc_j\omega^{jr}/D$ and $\bra{m}\XC\ket{n}=\XC_{mn}$ for general operator $\XC$, we have
\begin{align}\label{eqn27}
	\AC_{m_1m_2}\BC_{n_1n_2} = s_{m_1-m_2+d_1\left(n_1-n_2\right)},
\end{align}
for all $m_1,m_2=0,1,\cdots,d_1-1$ and $n_1,n_2=0,1,\cdots,D_2-1$. Since this depends only on $m_1-m_2$ and $n_1-n_2$, we can restrict consideration to cases where either $m_1=0$ or $m_2=0$ and $n_1=0$ or $n_2=0$. Now, considering the tautology $\left(\AC_{m_1m_2}\BC_{n_1n_2}\right)\left(\AC_{m_1^\prime m_2^\prime}\BC_{n_1^\prime n_2^\prime}\right)=\left(\AC_{m_1m_2}\BC_{n_1^\prime n_2^\prime}\right)\left(\AC_{m_1^\prime m_2^\prime}\BC_{n_1n_2}\right)$, we obtain from the restriction that $\RC$ is a product operator, that
\begin{align}\label{eqn28}
	s_{m_1+d_1n_1}s_{m_2+d_1n_2}=s_{m_1+d_1n_2}s_{m_2+d_1n_1},
\end{align}
for all $-d_1+1\le m_1,m_2\le d_1-1$ and $-D_2+1\le n_1,n_2\le D_2-1$.      %, and we lose nothing by assuming that $m_1\ne m_2$ and $n_1\ne n_2$.%Two special cases are when $m_1=0$ and when $n_1=0$.

Note that $s_0>0$ since $c_j\ge0$ for all $j$, and we can clearly assume $\RC\ne0$. Note also that $s_{-r}=s_r^\ast$ and $s_{r\pm N}=s_{r}$. Let us now prove the following lemma.
\begin{lem19}\label{lem19}
	If exists $q$ such that $s_q=0$, then $s_r=0$ for all $r\ne0$.
\end{lem19}
\proof To begin with, note that $\AC_{nn}\ne0$ and $\BC_{nn}\ne0$ because $0\ne s_0=\AC_{mm}\BC_{nn}$ for any $m,n$. Suppose $q=m_1+n_1d_1$ so that $0=s_{m_1+n_1d_1}=\AC_{m_10}\BC_{n_10}$, implying that either $\AC_{m_10}=0$ or $\BC_{n_10}=0$. First, suppose that $\AC_{m_10}=0$. Then, $0=\AC_{m_10}\BC_{01}=s_{m_1-d_1}=\left(s_{d_1-m_1}\right)^\ast$, implying $0=\AC_{d_1-m_1,0}\BC_{00}$. Since $\BC_{00}\ne0$, this means that $\AC_{d_1-m_1,0}=0$, and in turn, that $0=\AC_{d_1-m_1,0}\BC_{D_2-1,0}=s_{d_1-m_1+\left(D_2-1\right)d_1}$. Recall that $N=d_1D_2+1$ and that $\omega^N=1$. Therefore, the preceding expression becomes $0=s_{d_1-m_1+\left(D_2-1\right)d_1}=s_{N-m_1-1}=s_{-m_1-1}=\left(s_{m_1+1}\right)^\ast=\left(\AC_{m_1+1,0}\BC_{00}\right)^\ast$. Hence we see that $\AC_{m_10}=0\rightarrow \AC_{m_1+1,0}=0$, which means that $\AC_{m_10}=0\rightarrow \AC_{m0}=0$ for all $m\ge m_1$. In particular, we have that $d_1-1\ge m_1$ so that $0=\AC_{d_1-1,0}$, implying $0=\AC_{d_1-1,0}\BC_{01}=s_{-1}=\left(s_1\right)^\ast=\left(\AC_{10}\BC_{00}\right)^\ast$, so that in fact, $\AC_{10}=0$. Hence, for any fixed $m_1$ we have that $\AC_{m_10}=0$ implies $\AC_{10}=0$, which in turn implies that $\AC_{m0}=0$ for any $m$. Finally, this means that $s_{m+nd_1}=\AC_{m0}\BC_{n0}=0$ for any $m,n$ not both equal to zero, which proves the lemma in the first case that there exists $m_1$ such that $\AC_{m_10}=0$.

On the other hand if $\AC_{m_10}\ne0$, then $\BC_{n_10}=0$ when $s_{m_1+n_1d_1}=0$. This implies that $0=\BC_{n_10}\AC_{m0}=s_{m+n_1d_1}$ for any $m$. Then, $0=s_{m-d_1+\left(n_1+1\right)d_1}=\AC_{0,d_1-m}\BC_{n_1+1,0}$, so either (i) $\AC_{d_1-m,0}=0$ for some $m$, which by the preceding paragraph implies that $s_r=0$ for any $r\ne0$; or (ii) $\BC_{n_1+1,0}=0$. The latter case means that $\BC_{n_10}=0\rightarrow\BC_{n_1+1,0}=0\rightarrow\BC_{n0}=0$ for any $n\ge n_1$. Setting $n=D_2-1$, we have that $0=s_{\left(D_2-1\right)d_1}=s_{-1-d_1}=\left(s_{1+d_1}\right)^\ast=\left(\AC_{10}\BC_{10}\right)^\ast$, implying either (iia) $\AC_{10}=0\rightarrow s_r=0$ for any $r\ne0$, by the preceding paragraph; or (iib) $\BC_{10}=0$. The latter implies, by the foregoing argument, that $\BC_{n0}=0$ for any $n$, so that $s_{m+nd_1}=\AC_{m0}\BC_{n0}=0$ for any $m,n$, and the proof is complete.\endproof

Note that one can view $s_r$ as an inner product between vector $\vec c$ with components $c_j$ and vectors $\vec w_r$ with components $\omega^{jr}$, where the $N$ vectors $\vec w_r$ are easily seen to be mutually orthogonal. Hence, by this lemma, we have two possibilities: Either (1) $s_r=0$ for all $r\ne0$, in which case $\RC\propto I_A\otimes I_B$, which follows from \myeq{eqn23} and the fact that the only vector orthogonal to the $N-1$ vectors $\vec w_r$ with $r\ne0$ is $\vec w_0$, which has components that are independent of $j$; or (2) $s_r\ne0$ for all $r$. We now prove the following lemma.%show that in the latter case, rank$(\RC)=1$, and then in turn, that this implies $\RC\propto\Psi_j$ for some fixed $j$ \cmmt{another lemma for this?}.
\begin{lem20}\label{lem20}
	If $s_r\ne0$ for all $r$, then $rank\left(\RC\right)=1$, implying $\RC=c_j\Psi_j$ for some fixed $j$.
\end{lem20}
\proof With $\RC=\AC\otimes\BC=\sum_jc_j\Psi_j$ and $\Psi_j$ a rank-$1$ product operator, then $\BC\propto\sum_jc_j\textrm{Tr}_A(\Psi_j)$ is a positive linear combination of rank-$1$ positive operators. Since rank-$1$ positive operators are extreme rays in the convex cone of positive operators, then if $\textrm{rank}\left(\BC\right)=1$, it must be that there is one and only one non-zero $c_j$, which implies both that $\textrm{rank}\left(\RC\right)=1$ and that $\RC=c_j\Psi_j$ for some fixed $j$. Therefore, we need only show that $\textrm{rank}\left(\BC\right)=1$. This will be so if every $2\times2$ submatrix of $\BC$ has determinant equal to zero, or in other words, if $\BC_{n_1n_2}\BC_{n_3n_4}=\BC_{n_1n_4}\BC_{n_3n_2}$ for all $n_1,n_2,n_3,n_4$. Within each of these $2\times2$ submatrices, we may choose to call the lower-right element as $\BC_{n_1n_2}$, and then without loss of generality, we have that $n_1>n_3\ge0$ and $n_2>n_4\ge0$. These conditions on $\BC$ are equivalent to,
\begin{align}\label{eqn29}
	s_{\left(n_1-n_2\right)d_1}s_{\left(n_3-n_4\right)d_1}=s_{\left(n_1-n_4\right)d_1}s_{\left(n_3-n_2\right)d_1},
\end{align}
for all $n_1>n_3\ge0$ and $n_2>n_4\ge0$, so these are what we need to show follow from \myeq{eqn28}.

We have
\begin{align}\label{eqn30}
	s_{\left(n_1-n_2\right)d_1}s_{\left(n_3-n_4\right)d_1}&=s_{\left(n_1-n_2\right)d_1}s_{\pm1+\left(n_3-n_4\pm D_2\right)d_1}\notag\\
	&=s_{\pm1+\left(n_1-n_2\right)d_1}s_{\left(n_3-n_4\pm D_2\right)d_1}\notag\\
	&=s_{\mp d_1\pm1+\left(n_1-n_2\pm1\right)d_1}s_{\left(n_3-n_4\pm D_2\right)d_1}\notag\\
	&=s_{\left(n_1-n_2\pm1\right)d_1}s_{\mp d_1\pm1+\left(n_3-n_4\pm D_2\right)d_1}\notag\\
	&=s_{\left(n_1-n_2\pm1\right)d_1}s_{\left(n_3-n_4\mp1\right)d_1}\notag\\
\end{align}
where we have used \myeq{eqn28} along with the fact that $s_{r\pm N}=s_r$, where $N=d_1D_2+1$. The upper sign must be chosen if $n_3<n_4$, the lower sign if $n_3>n_4$. For the upper sign, we repeat this process $n_2-n_4$ times, while for the lower sign we repeat it $n_1-n_3$ times. In either case, we end up with \myeq{eqn29}, as desired. When $n_3=n_4$ we can swap the roles of the pair $n_3,n_4$ with the pair $n_1,n_2$ and obtain the desired result, except when it is also the case that $n_1=n_2$.\footnote{Note that a problem arises with this process if we ever end up with $s_{D_2d_1}$ or $s_{-D_2d_1}$, since then \myeq{eqn28} doesn't apply. However, $s_{D_2d_1}$ can only appear after repeating the process $ n=D_2-n_1+n_2$ times for the upper sign, or for $ n=D_2-n_3+n_4$ for the lower sign. For the upper sign this would only happen if $n_2-n_4\ge D_2-n_1+n_2$, or if $n_1-n_4\ge D_2$, which is a contradiction, and for the lower sign it only happens if $n_1-n_3\ge D_2-n_3+n_4$, giving the same contradiction. On the other hand, $s_{-D_2d_1}$ can only appear after repeating the process $ n=D_2+n_3-n_4$ times for the upper sign, or for $ n=D_2+n_1-n_2$ for the lower sign. For the upper sign, this requires that $n_2-n_4\ge D_2+n_3-n_4$, and for the lower sign it requires that $n_1-n_3\ge D_2+n_1-n_2$. Both of these lead again to a contradiction, in this case $n_2-n_3\ge D_2$, so this problem does not arise.} Therefore, we are done except for showing that $s_0^2=s_qs_{-q}$ for all $q$ in the range $0<q\le d_1\left(D_2-1\right)$.

From \myeq{eqn28}, we have that $s_0s_{m_2+d_1n_2}=s_{d_1n_2}s_{m_2}$. Multiplying this by $s_0$, we obtain
\begin{align}\label{eqn31}
	s_0^2s_{m_2+d_1n_2}=s_0s_{d_1n_2}s_{m_2}=s_0\left[s_{-1-\left(D_2-n_2\right)d_1}s_{m_2}\right]=s_{-1}s_{-\left(D_2-n_2\right)d_1}s_{m_2}=s_{-1}\left[s_{1+n_2d_1}s_{m_2}\right]=s_{-1}s_1s_{m_2+n_2d_1}.
\end{align}
Since by assumption, $s_r\ne0$ for all $r$, this implies that
\begin{align}\label{eqn32}
	s_0^2=s_1s_{-1}=s_1s_{d_1-1-d_1}=s_{d_1-1}s_{1-d_1}=s_{d_1-1}s_{2+\left(D_2-1\right)d_1}=s_2s_{-1+D_2d_1}=s_2s_{-2}=s_2s_{d_1-2-d_1}=s_{d_1-2}s_{2-d_1},
\end{align}
and so on, where we have repeatedly used \myeq{eqn28} and the fact that $s_r=s_{r\pm N}=s_{r\pm\left(D_2d_1+1\right)}$. This shows that $s_0^2=s_rs_{-r}$ for all $r=0,1,\cdots,d_1-1$. From here, we find
\begin{align}\label{eqn33}
	s_0^2=s_{-1+d_1}s_{1-d_1}=s_{1+d_1}s_{-1-d_1}=s_{1+d_1}s_{\left(D_2-1\right)d_1}=s_{d_1}s_{1+\left(D_2-1\right)d_1}=s_{d_1}s_{-d_1},
\end{align}
and then starting from
\begin{align}\label{eqn34}
	s_0^2=s_{1+nd_1}s_{-1-nd_1},
\end{align}
we find that
\begin{align}\label{eqn35}
	s_0^2&=s_{1-d_1+\left(n+1\right)d_1}s_{-1-nd_1}=s_{-1+\left(n+1\right)nd_1}s_{1-d_1-nd_1}=s_{1+\left(n+1\right)d_1}s_{-1-\left(n+1\right)d_1},
\end{align}
from which we have that $s_0^2=s_{1+nd_1}s_{-1-nd_1}$ for all $n$. From this, we find that
\begin{align}\label{eqn36}
	s_0^2&=s_{1+nd_1}s_{-1-nd_1}=s_{1+nd_1}s_{\left(D_2-n\right)d_1}=s_{nd_1}s_{1+\left(D_2-n\right)d_1}=s_{nd_1}s_{-nd_1},
\end{align}
which tells us that $s_0^2=s_{nd_1}s_{-nd_1}$ for every $n$. Finally, by following the steps taken in \myeq{eqn32} but instead starting from \myeq{eqn34}, we have that
\begin{align}\label{eqn37}
	s_0^2=s_{-1+nd_1}s_{1-nd_1}=s_{-1+nd_1}s_{2+\left(D_2-n\right)d_1}=s_{2+nd_1}s_{-1+\left(D_2-n\right)d_1}=s_{2+nd_1}s_{-2-nd_1}=\cdots,
\end{align}
and thus we have that $s_0^2=s_{m+nd_1}s_{-m-nd_1}$ for every $m,n$, which completes the proof.\endproof

We thus have two possibilities for $\RC=\sum_jc_j\Psi_j$ to be a product operator. Either (1) $\RC\propto I$, the identity operator on the full Hilbert space $\HC$, or (2) $\RC\propto\Psi_j$ for some fixed $j$. As a consequence, there cannot exist a continuous path of product operators stretching from $I$ to any one of the $\Psi_j$ in the space of positive operators on $\HC$. By Corollary~\ref{cor1}, this completes the proof of Theorem~\ref{thm11}.

%\bibliography{Qrefs}
%\bibliographystyle{prsty}

\end{document}